\documentclass[10pt,journal,twoside]{IEEEtran}

\usepackage[cmex10]{amsmath}
\usepackage[caption=false,font=footnotesize]{subfig}
\usepackage{graphicx, amssymb, xcolor, cite, booktabs, multirow, url, pgfplots, bm, bbm}

\newtheorem{theorem}{Theorem}

\newtheorem{proposition}{Proposition}
\newtheorem{lemma}{Lemma}
\newtheorem{corollary}{Corollary}
\newtheorem{conjecture}{Conjecture}

\newcommand{\eps}{\varepsilon}
\newcommand{\var}{\operatorname{Var}}
\newcommand{\expn}{\mathbb{E}}
\newcommand{\pr}{\mathbb{P}}
\newcommand{\pmark}{\mathcal{P}_{\text{mark}}}
\renewcommand{\vec}{\bm}

\newcommand{\thint}{\vec{\theta}_{\text{int}}}
\newcommand{\thall}{\vec{\theta}_{\text{all1}}}
\newcommand{\thmaj}{\vec{\theta}_{\text{maj}}}
\newcommand{\thmin}{\vec{\theta}_{\text{min}}}
\newcommand{\thcoi}{\vec{\theta}_{\text{coin}}}
\newcommand{\thadd}{\vec{\theta}_{\text{add}}}
\newcommand{\thdil}{\vec{\theta}_{\text{dil}}}

\newcommand{\tos}{^s}

\newcommand{\palls}{p_{\text{all1}}\tos}

\begin{document}

%\markboth{IEEE Transactions on Information Theory, Vol.~1, No.~1, January~2000}{Laarhoven: Asymptotics of Fingerprinting and Group Testing: Decoders}

\title{Asymptotics of Fingerprinting and Group Testing: Capacity-Achieving Log-Likelihood Decoders}
\author{Thijs~Laarhoven%
\thanks{T.~Laarhoven is with the Department of Mathematics and Computer Science, Eindhoven University of Technology, The Netherlands.\protect\\
E-mail: {\ttfamily mail@thijs.com}.\protect\\%
Part of the material in this paper will be presented at the 2nd ACM Workshop on Information Hiding and Multimedia Security (Salzburg, Austria, June 2014).}}

\maketitle

\begin{abstract}
We study the large-coalition asymptotics of fingerprinting and group testing, and derive explicit decoders that provably achieve capacity for many of the considered models. We do this both for simple decoders (fast but suboptimal) and for joint decoders (slow but optimal), and both for informed and uninformed settings.

For fingerprinting, we show that if the pirate strategy is known, the Neyman-Pearson-based log-likelihood decoders provably achieve capacity, regardless of the strategy. The decoder built against the interleaving attack is further shown to be a universal decoder, able to deal with arbitrary attacks and achieving the uninformed capacity. This universal decoder is shown to be closely related to the Lagrange-optimized decoder of Oosterwijk et al.\ and the empirical mutual information decoder of Moulin. Joint decoders are also proposed, and we conjecture that these also achieve the corresponding joint capacities. 

For group testing, the simple decoder for the classical model is shown to be more efficient than the one of Chan et al.\ and it provably achieves the simple group testing capacity. For generalizations of this model such as noisy group testing, the resulting simple decoders also achieve the corresponding simple capacities.
\end{abstract}


\IEEEpeerreviewmaketitle

\section{Introduction}
\label{sec:intro}

%WWWWWWWWWWWWWWWWWWWWWWWWWWWWWWWWWWWWWWWWWWWWWWWWWWWWWWWW

\subsection{Fingerprinting}
\label{sec:intro-fp}

\IEEEPARstart{T}{o} protect copyrighted content against unauthorized redistribution, distributors commonly embed watermarks or fingerprints in the content, uniquely linking copies to individual users. If the distributor finds an illegal copy of the content online, he can then extract the watermark from this copy and compare it to the database of watermarks, to determine which user was responsible. 

To combat this solution, a group of $c$ pirates may try to form a coalition and perform a collusion attack. By comparing their unique versions of the content, they will detect differences in their copies which must be part of the watermark. They can then try to create a mixed pirate copy, where the resulting watermark matches the watermark of different pirates in different segments of the content, making it hard for the distributor to find the responsible users. The goal of the distributor of the content is to assign the watermarks to the users in such a way that, even if many pirates collude, the pirate copy can still be traced back to the responsible users.

%WWWWWWWWWWWWWWWWWWWWWWWWWWWWWWWWWWWWWWWWWWWWWWWWWWWWWWWW

\subsection{Group testing}
\label{sec:intro-gt}

A different area of research that has received considerable attention in the last few decades is group testing, introduced by Dorfman~\cite{dorfman43} in the 1940s. Suppose a large population contains a small number $c$ of infected (or defective) items. To identify these items, it is possible to perform group tests: testing a subset of the population will lead to a positive test result if this subset contains at least one defective item, and a negative result otherwise. Since the time to run a single test may be very long, the subsets to test need to be chosen in advance, after which all group tests are performed simultaneously. Then, when the test results come back, the subset of defective items needs to be identified. The goal of the game is to identify these defectives using as few group tests as possible, and with a probability of error as small as possible. 

%WWWWWWWWWWWWWWWWWWWWWWWWWWWWWWWWWWWWWWWWWWWWWWWWWWWWWWWW

\subsection{Model}
\label{sec:intro-model}

The above problems of fingerprinting and group testing can be jointly modeled by the following two-person game between (in terms of fingerprinting) the distributor $\mathcal{D}$ and the adversary $\mathcal{C}$ (the set of colluders, or the set of defectives). Throughout the paper we will mostly use terminology from fingerprinting (i.e. users instead of items, colluders instead of defective items), unless we are specifically dealing with group testing results.

First, there is a universe $\mathcal{U}$ of $n$ users, and the adversary is assigned a random subset of users $\mathcal{C} \subseteq \mathcal{U}$ of size $|\mathcal{C}| = c$. This subset $\mathcal{C}$ is unknown to the distributor (but we assume that the distributor does know the size $c$ of $\mathcal{C}$), and the aim of the game for the distributor is ultimately to discover $\mathcal{C}$. The two-person game consists of three phases: (1) the distributor uses an \textit{encoder} to generate a fingerprinting code, used for assigning versions to users; (2) the colluders employ a \textit{collusion channel} to generate the pirate output from their given code words; and (3) the distributor uses a \textit{decoder} to map the pirate output to a set $\mathcal{C}' \subseteq \mathcal{U}$.

%-------------------------------------------------------%

\subsubsection{Encoder}

The distributor generates a fingerprinting code $\mathcal{X}$ of $n$ binary code words of length $\ell$.\footnote{In fingerprinting a common generalization is to assume that the entries of the code words come from an alphabet of size $q \geq 2$, but in this paper we restrict our attention to the binary case $q = 2$.} The parameter $\ell$ is referred to as the code length, and the distributor would like $\ell$ to be as small as possible. For the eventual embedded watermark, we assume that for each segment of the content there are two differently watermarked versions, so the watermark of user $j$ is determined by the $\ell$ entries in the $j$th code word of $\mathcal{X}$. 

A common restriction on the encoding process is to assume that $\mathcal{X}$ is created by first generating a bias vector $\vec{P} \in (0,1)^{\ell}$ (by choosing each entry $P_i$ independently from a certain distribution $f_P$), and then generating code words $\vec{X}_j \in \mathcal{X}$ according to $\pr(X_{j,i} = 1) = P_i$. This guarantees that watermarks of different users $j$ are independent, and that watermarks in different positions $i$ are independent. Fingerprinting schemes that satisfy this assumption are sometimes called bias-based schemes, and the encoders in this paper (both for group testing and fingerprinting) are also assumed to belong to this category.

%-------------------------------------------------------%

\subsubsection{Collusion channel}

After generating $\mathcal{X}$, the code words are used to select and embed watermarks in the content, and the content is sent out to all users. The colluders then get together, compare their copies, and use a certain collusion channel or pirate attack $\vec{\Theta}$ to determine the pirate output $\vec{Y} \in \{0,1\}^{\ell}$. If the pirate attack behaves symmetrically both in the colluders and in the positions $i$, then the collusion channel can be modeled by a vector $\vec{\theta} \in [0,1]^{c+1}$, consisting of entries $\theta_z = f_{Y_i|Z_i}(1|z) = \pr(Y_i = 1|Z = z)$ (for $z = 0, \dots, c$) indicating the probability of outputting a $1$ when the pirates received $z$ ones and $c - z$ zeroes. A further restriction on $\vec{\theta}$ in fingerprinting is the marking assumption, introduced by Boneh and Shaw~\cite{boneh98}, which says that $\theta_0 = 0$ and $\theta_c = 1$, i.e., if the pirates receive only zeros or ones they have to output this symbol.

%-------------------------------------------------------%

\subsubsection{Decoder}

Finally, after the pirate output has been generated and distributed, we assume the distributor intercepts it and applies a decoding algorithm to $\vec{Y}$, $\mathcal{X}$ and $\vec{P}$ to compute a set $\mathcal{C}' \subseteq \mathcal{U}$ of accused users. The distributor wins the game if $\mathcal{C}' = \mathcal{C}$ and loses if this is not the case.

%-------------------------------------------------------%

\subsubsection*{Fingerprinting vs. group testing}

While the above model is described in fingerprinting terminology, it also covers many common group testing models. The users then correspond to items, the colluders translate to defectives, the code $\mathcal{X}$ corresponds to the group testing matrix $X$ (where $X_{j,i} = 1$ if item $j$ is included in the $i$th test), and the pirate output corresponds to positive/negative test results. The collusion channel is exactly what separates group testing from fingerprinting: while in fingerprinting it is commonly assumed that this channel is not known or only weakly known to the distributor, in group testing this channel is usually assumed known in advance. This means that there is no malicious adversary in group testing, but only a randomization procedure that determines $\vec{Y}$. Note also that in (noisy) group testing, the Boneh-Shaw marking assumption may not always hold.

%WWWWWWWWWWWWWWWWWWWWWWWWWWWWWWWWWWWWWWWWWWWWWWWWWWWWWWWW

\subsection{Related work}
\label{sec:intro-related}

Work on the fingerprinting game described above started in the late 90s, and lower bounds on the code length were established of the order $\ell \propto c \ln n$~\cite{boneh98}, until in 2003 Tardos~\cite{tardos03} proved a lower bound of the order $\ell \propto c^2 \ln n$ and described a scheme with $\ell = O(c^2 \ln n)$, showing this bound is tight. Since the leading constants of the upper and lower bounds did not match, later work on fingerprinting focused on finding the optimal leading constant. 

Based on channel capacities, Amiri and Tardos~\cite{amiri09} and Huang and Moulin~\cite{huang12} independently derived the optimal leading constant to be $2$ (i.e.\ an asymptotic code length of $\ell \sim 2 c^2 \ln n$) and many improvements to Tardos's original scheme were made~\cite{blayer08, laarhoven14dcc, nuida09, skoric08} to reduce the leading constant from $100$ to $\frac{1}{2} \pi^2 \approx 4.93$. Recently it was shown that with Tardos' original `score function' one cannot achieve capacity~\cite{laarhoven13ihmmsec}, which lead to the study of different score functions. Based on a result of Abbe and Zheng~\cite{abbe10}, Meerwald and Furon~\cite{meerwald12} noted that a score function designed against the worst-case attack achieves capacity against arbitrary attacks. This also lead to a proposal for a capacity-achieving score function in~\cite{oosterwijk13b}, which achieves the lower bound on the leading constant of $2$.

Most of the work on fingerprinting focused on the setting of arbitrary, unknown attacks, but some work was also done on the informed setting, where the decoder knows or tries to estimate the pirate strategy~\cite{charpentier09, furon09b, laarhoven13wifs, meerwald11b, oosterwijk13b}. It is well known that for suboptimal pirate attacks the required code length may be significantly smaller than $\ell \sim 2 c^2 \ln n$, but explicit schemes provably achieving an optimal scaling in $\ell$ are not known.

Research on the group testing problem started much longer ago, and already in 1985 exact asymptotics on the code length for probabilistic schemes were derived as $\ell \sim c \log_2 n$~\cite{sebo85}, whereas deterministic schemes require a code length of $\ell \propto c^2 \ln n$~\cite{dyachkov82, dyachkov89}. Later work focused on slight variations of the classical model such as noisy group testing, where a positive result may not always correspond to the presence of a defective item due to `noise' in the test output~\cite{atia12, cheraghchi11, laarhoven13allerton, sejdinovic10}. For noisy group testing exact asymptotics on the capacities (with leading constants) are yet unknown, and so it is not known whether existing constructions are optimal.

%WWWWWWWWWWWWWWWWWWWWWWWWWWWWWWWWWWWWWWWWWWWWWWWWWWWWWWWW

\subsection{Contributions and outline}
\label{sec:intro-contributions}

In this paper we show how to build schemes using log-likelihood decoders that provably satisfy given bounds on the error probabilities, and have a code length with the optimal asymptotic scaling. We do this both for the informed setting (where $\vec{\theta}$ is known to the decoder) and the universal setting, where $\vec{\theta}$ is only known to satisfy the marking assumption. The results for the informed setting for fingerprinting are summarized in Table~\ref{tab:tab1}. Simple informed and universal decoders are discussed in Sections~\ref{sec:dec-simple} and \ref{sec:dec-simple-universal}, and joint informed and universal decoders are discussed in Sections~\ref{sec:dec-joint} and \ref{sec:dec-joint-universal} respectively. In-between, Section~\ref{sec:dec-simple-universal-emi} discusses a completely different approach to obtain a universal simple decoder (based on the empirical mutual information decoder of Moulin~\cite{moulin08}) and shows how in the end the result is again quite similar.

\begin{table}[!t]
\renewcommand{\arraystretch}{1.3}
\caption{An overview of the provable asymptotic code lengths of the informed fingerprinting decoders discussed in this paper. \label{tab:tab1}}
\centering
\begin{tabular}{lll} \toprule
Fingerprinting attack & Simple decoding & Joint decoding \\ \midrule
%Fingerprinting & & & & & \\
$\thint$: interleaving attack & $\ell \sim 2 c^2 \ln n$ & $\ell \sim 2c^2 \ln n$ \\
$\thall$: all-$1$ attack & $\ell \sim \frac{c \ln n}{(\ln 2)^2}$ & $\ell \sim c \log_2 n$ \\
$\thmaj$: majority voting & $\ell \sim \pi c \ln n$ & $\ell \sim c \log_2 n$ \\
$\thmin$: minority voting & $\ell \sim \frac{c \ln n}{(\ln 2)^2}$ & $\ell \sim c \log_2 n$ \\
$\thcoi$: coin-flip attack & $\ell \sim \frac{4 c \ln n}{(\ln 2)^2}$ & $\ell \sim c \log_{5/4} n$ \\ \bottomrule
%\parbox[t]{2mm}{\multirow{6}{*}{\rotatebox[origin=c]{90}{Group testing}}} & $\thall$: classical model & $\frac{c \ln n}{(\ln 2)^2}$ & $c \log_2 n$ \\
%& $\thadd$: additive noise & $\frac{c \ln n}{(\ln 2)^2}(1 + O(r))$ & $c \log_2 n (1 + O(r \ln r))$ \\
%& $\thdil$: dilution noise & $\frac{c \ln n}{(\ln 2)^2}(1 + O(r \ln r))$ & $c \log_2 n (1 + O(r \ln r))$ \\ \bottomrule
\end{tabular}
\end{table}

%WWWWWWWWWWWWWWWWWWWWWWWWWWWWWWWWWWWWWWWWWWWWWWWWWWWWWWWW
%WWWWWWWWWWWWWWWWWWWWWWWWWWWWWWWWWWWWWWWWWWWWWWWWWWWWWWWW
%WWWWWWWWWWWWWWWWWWWWWWWWWWWWWWWWWWWWWWWWWWWWWWWWWWWWWWWW
%WWWWWWWWWWWWWWWWWWWWWWWWWWWWWWWWWWWWWWWWWWWWWWWWWWWWWWWW

\section{Simple informed decoding}
\label{sec:dec-simple}

In this section we will discuss simple decoders with explicit scheme parameters (code lengths, accusation thresholds) that provably satisfy given bounds on the error probabilities. The asymptotics of the resulting code lengths further show that these schemes are capacity-achieving; asymptotically, the code lengths achieve the lower bounds that follow from the simple capacities, as derived in~\cite{laarhoven14capacities}. 

We will follow the bias-based and score-based framework introduced by Tardos~\cite{tardos03}, which was later generalized to joint decoders by Moulin~\cite{moulin08}. For simple decoding, this means that a user $j$ receives a score $S_j$ of the form
\begin{align}
S_j = \sum_{i=1}^{\ell} S_{j,i} = \sum_{i=1}^{\ell} g(x_{j,i}, y_i, p_i)
\end{align}
and he is accused iff $S_j \geq \eta$ for some fixed threshold $\eta$. The function $g$ is sometimes called the score function. Note that since $g$ only depends on $\mathcal{X}$ through $\vec{X}_j$, any decoder that follows this framework is a simple decoder. 

%WWWWWWWWWWWWWWWWWWWWWWWWWWWWWWWWWWWWWWWWWWWWWWWWWWWWWWWW

\subsection{Simple log-likelihood decoders}
\label{sec:dec-simple-intro}

Several different score functions $g$ have been considered before~\cite{laarhoven13wifs, oosterwijk13b, skoric08, tardos03}, but in this work we will restrict our attention to log-likelihood scores, which are known to perform well and which turn out to be quite easy to analyze. 

First, when building a decoder we naturally want to be able to distinguish between two cases: user $j$ is guilty or user $j$ is not guilty. To do this, we assign scores to users based on the available data, and we try to obtain an optimal trade-off between the false positive error (accusing an innocent user) and the false negative error (not accusing a guilty user). This problem is well known in statistics as a hypothesis testing problem, where in this case we want to distinguish between the following two hypotheses $H_0$ and $H_1$:
\begin{align}
&H_0: \quad \text{user $j$ is guilty } (j \in \mathcal{C}), \\
&H_1: \quad \text{user $j$ is innocent } (j \notin \mathcal{C}).
\end{align}
The Neyman-Pearson lemma~\cite{neyman33} tells us that the most powerful test to distinguish between $H_0$ and $H_1$ is to test whether the following likelihood ratio exceeds an appropriately chosen threshold $\eta$:
\begin{align}
\Lambda(\vec{x}, \vec{y}, \vec{p}) = \frac{f_{\vec{X},\vec{Y}|\vec{P}}(\vec{x},\vec{y}|\vec{p},H_0)}{f_{\vec{X},\vec{Y}|\vec{P}}(\vec{x},\vec{y}|\vec{p},H_1)}.
\end{align}
Taking logarithms, and noting that different positions $i$ are i.i.d., it is clear that testing whether a user's likelihood ratio exceeds $\eta_1$ is equivalent to testing whether his score $S_j$ exceeds $\eta = \ln \eta_1$ for $g$ defined by
\begin{align}
g(x,y,p) = \ln\left(\frac{f_{X,Y|P}(x,y|p,H_0)}{f_{X,Y|P}(x,y|p,H_1)}\right). \label{eq:dec-simple-g}
\end{align}
Thus, the score function $g$ from~\eqref{eq:dec-simple-g} corresponds to using a Neyman-Pearson score over the entire code word $\vec{X}_j$, and therefore $g$ is in a sense optimal for minimizing the false positive error for a fixed false negative error. Score functions of this form were previously considered in the context of fingerprinting in e.g.\ \cite{meerwald11b, perez09}, but these papers did not show how to choose $\eta$ and $\ell$ to provably satisfy certain bounds on the error probabilities.

%WWWWWWWWWWWWWWWWWWWWWWWWWWWWWWWWWWWWWWWWWWWWWWWWWWWWWWWW

\subsection{Theoretical evaluation}
\label{sec:dec-simple-theory}

Let us first see how we can choose $\ell$ and $\eta$ such that we can prove that the false positive and false negative error probabilities are bounded from above by certain values $\eps_1$ and $\eps_2$. For the analysis below, we will make use of the following function $M$, which is closely related to the moment-generating function of scores in one position $i$ for both innocent and guilty users. For fixed $p$, this function $M$ is defined on $[0,1]$ by
\begin{align}
M(t) = \sum_{x,y} f_{X,Y|P}(x,y|p,H_0)^t f_{X,Y|P}(x,y|p,H_1)^{1-t}, \label{eq:M}
\end{align}
and it satisfies $M(t) = \expn(e^{t S_{j,i}}|p,H_1) = \expn(e^{(t - 1)S_{j,i}}|p,H_0)$. 

\begin{theorem} \label{thm:dec-simple-informed}
Let $p$ and $\vec{\theta}$ be fixed and known to the decoder. Let $\gamma = \ln(1/\eps_2) / \ln(n/\eps_1)$, and let the code length $\ell$ and threshold $\eta$ be chosen as
\begin{align}
\ell &= \frac{\sqrt{\gamma} (1 + \sqrt{\gamma} - \gamma)}{-\ln M(1 - \sqrt{\gamma})} \, \ln\left(\frac{n}{\eps_1}\right), \label{eq:dec-simple-ell} \\ 
\eta &= (1 - \gamma)\ln\left(\frac{n}{\eps_1}\right). \label{eq:dec-simple-eta}
\end{align}
Then with probability at least $1 - \eps_1$ no innocent users are accused, and with probability at least $1 - \eps_2$ at least one colluder is caught.
\end{theorem}

\begin{IEEEproof}
For innocent users $j$, we would like to prove that $\pr(S_j > \eta | H_0) \leq \frac{\eps_1}{n}$, where $S_j$ is the user's total score over all segments. If this can be proved, then since innocent users have independent scores, it follows that with probability at least $(1 - \frac{\eps_1}{n})^n \geq 1 - \eps_1$ no innocent users are accused. To get somewhat tight bounds, we start by applying the Markov inequality to $e^{\alpha S_j}$ for some $\alpha > 0$:
\begin{align}
\pr(S_j > \eta | H_1) = \pr(e^{\alpha S_j} > e^{\alpha \eta}|H_1) \leq \frac{\expn(e^{\alpha S_j} | H_1)}{e^{\alpha \eta}} \\
= \frac{\prod_{i=1}^{\ell} \expn(e^{\alpha S_{j,i}} | H_1)}{e^{\alpha \eta}} = \frac{M(\alpha)^{\ell}}{e^{\alpha \eta}}.
\end{align}
Let us now first switch to guilty users. We will prove that for some guilty user $j$, $\pr(S_j < \eta | H_1) \leq \eps_2$. Again using Markov's inequality with some fixed constant $\beta > 0$, we get
\begin{align}
\pr(S_j < \eta | H_0) = \pr(e^{-\beta S_j} > e^{-\beta \eta}|H_0) \leq \frac{\expn(e^{-\beta S_j} | H_0)}{e^{-\beta \eta}} \\
= \frac{\prod_{i=1}^{\ell} \expn(e^{-\beta S_{j,i}} | H_0)}{e^{-\beta \eta}} = \frac{M(1 - \beta)^{\ell}}{e^{-\beta \eta}}.
\end{align}
For both guilty and innocent users, we can now obtain bounds by choosing appropriate values for $\alpha$ and $\beta$. Investigating the resulting expressions, it seems that good choices for $\alpha, \beta$ leading to sharp bounds are $\alpha = 1 - \sqrt{\gamma}$ and $\beta = \sqrt{\gamma}$. Substituting these choices for $\alpha$ and $\beta$, and setting the bounds equal to the desired upper bounds $\frac{\eps_1}{n}$ and $\eps_2$, we get
\begin{align}
\pr(S_j > \eta | H_1) &\leq \frac{M(1 - \sqrt{\gamma})^{\ell}}{e^{-\sqrt{\gamma}\eta}} \, e^{-\eta} = \frac{\eps_1}{n}\, , \\
\pr(S_j < \eta | H_0) &\leq \frac{M(1 - \sqrt{\gamma})^{\ell}}{e^{-\sqrt{\gamma}\eta}} = \eps_2.
\end{align}
Combining these equations we obtain the given expression for $\eta$, and solving for $\ell$ leads to the expression for $\ell$ in~\eqref{eq:dec-simple-ell}.
\end{IEEEproof}

Compared to previous papers analyzing provable bounds on the error probabilities~\cite{blayer08, ibrahimi13, laarhoven14dcc, skoric08, tardos03}, the proof of Theorem~\ref{thm:dec-simple-informed} is remarkably short and simple. Note however that the proof that a colluder is caught assumes that the attack used by the colluders is the same as the one the decoder is built against, and that the actual value of $\ell$ is still somewhat mysterious due to the term $M(1 - \sqrt{\gamma})$. In Section~\ref{sec:dec-simple-codelength} we will show how to get some insight into this expression for $\ell$.

%WWWWWWWWWWWWWWWWWWWWWWWWWWWWWWWWWWWWWWWWWWWWWWWWWWWWWWWW

\subsection{Practical evaluation}
\label{sec:dec-simple-practice}

Before going into details how the code lengths of Theorem~\ref{thm:dec-simple-informed} scale, note that Theorem~\ref{thm:dec-simple-informed} only shows that with high probability we provably catch \textit{at least one} colluder with this decoder. Although this is commonly the best you can hope for when dealing with arbitrary attacks in fingerprinting\footnote{In those cases attacks exist guaranteeing you will not catch more than one colluder, such as the `scapegoat' strategy~\cite{laarhoven13tit}.}, if the attack is colluder-symmetric it is actually possible to catch \textit{all} colluders with high probability. So instead we would like to be able to claim that with high probability, the set of accused users $\mathcal{C}'$ \textit{equals} the set of colluders $\mathcal{C}$. Similar to the proof for innocent users, we could simply replace $\eps_2$ by $\frac{\eps_2}{c}$, and argue that the probability of finding all pirates is the product of their individual probabilities of getting caught, leading to a lower bound on the success probability of $(1 - \frac{\eps_2}{c})^c \geq 1 - \eps_2$. This leads to the following heuristic estimate for the code length required to catch \textit{all} pirates.

\begin{conjecture}
Let $\gamma$ in Theorem~\ref{thm:dec-simple-informed} be replaced by $\gamma' = \ln(c/\eps_2) / \ln(n/\eps_1)$. Then with probability at least $1 - \eps_1$ no innocent users are accused, and with probability at least $1 - \eps_2$ \textit{all colluders are caught}.
\end{conjecture}

The problem with this claim is that the pirate scores are related through $\vec{Y}$, so they are not independent. As a result, we cannot simply take the product of the individual probabilities $(1 - \frac{\eps_2}{c})$ to get a lower bound on the success probability of $1 - \eps_2$. On the other hand, especially when the code length $\ell$ is large and $\eps_2$ is small, we do not expect the event $\{S_1 > T\}$ to tell us much about the probability of e.g.\ $\{S_2 > T\}$ occurring. One might thus expect that $\{S_2 > T\}$ does not become much less likely when $\{S_1 > T\}$ occurs. But since it is not so simple to prove a rigorous upper bound on the \textit{catch-all} error probability without assuming independence, we leave this problem for future work.

%WWWWWWWWWWWWWWWWWWWWWWWWWWWWWWWWWWWWWWWWWWWWWWWWWWWWWWWW

\subsection{Asymptotic code lengths}
\label{sec:dec-simple-codelength}

Let us now see how the code lengths $\ell$ from~\eqref{eq:dec-simple-ell} scale in terms of $c$ and $n$. In general this expression is not so pretty, but if we focus on the regime of large $n$ (and fixed $\eps_1$ and $\eps_2$), it turns out that the code length always has the optimal asymptotic scaling, regardless of $p$ and $\vec{\theta}$.

\begin{theorem} \label{thm:dec-simple-asymptotics}
For large $n$ and fixed $\eps_1$ and $\eps_2$, the code length $\ell$ of Theorem~\ref{thm:dec-simple-informed} scales as
\begin{align}
\ell = \frac{\log_2 n}{I(X_1;Y|P = p)}[1 + O(\sqrt{\gamma})], \label{eq:dec-simple-asymptotics}
\end{align}
where $I(X_1;Y|P = p)$ is the mutual information between a pirate symbol $X_1$ and the pirate output $Y$. As a result, $\ell$ has the optimal asymptotic scaling.
\end{theorem}

\begin{IEEEproof}
First, note that if $n \to \infty$ and $\eps_1, \eps_2$ are fixed, then $\gamma \to 0$. Let us first study the behavior of $M(1 - \sqrt{\gamma})$ for small $\gamma$, by computing the first order Taylor expansion of $M(1 - \sqrt{\gamma})$ around $\gamma = 0$. For convenience, below we will abbreviate $f_{X,Y|P}(x,y|p)$ by $f(x,y|p)$.
\begin{align}
& \quad M(1 - \sqrt{\gamma}) \nonumber \\
 &= \sum_{x,y} f(x,y|p,H_0) \exp\left(-\sqrt{\gamma} \ln\left(\frac{f(x,y|p,H_0)}{f(x,y|p,H_1)}\right)\right) \\
 &= \sum_{x,y} f(x,y|p,H_0) \left(1 - \sqrt{\gamma} \ln\left(\frac{f(x,y|p,H_0)}{f(x,y|p,H_1)}\right) + O(\gamma)\right) \\
 &= 1 - \sqrt{\gamma} \sum_{x,y} f(x,y|p,H_0) \ln\left(\frac{f(x,y|p,H_0)}{f(x,y|p,H_1)}\right) + O(\gamma).
\end{align}
Here the second equality follows from the fact that if $f(x,y|p, H_0) = 0$, then the factor $f(x,y|p,H_0)$ in front of the exponentiation would already cause this term to be $0$, while if $f(x,y|p,H_0) > 0$, then also $f(x,y|p,H_1) > 0$ and thus their ratio is bounded and does not depend on $\gamma$. Now, recognizing the remaining summation as the mutual information (in natural units) between a colluder symbol $X_1$ and the pirate output $Y$, we finally obtain:
\begin{align}
M(1 - \sqrt{\gamma}) = 1 - I(X_1;Y|P = p) \sqrt{\gamma} \ln 2 + O(\gamma).
\end{align}
Substituting this result in the original equation for $\ell$, and noting that the factor $\eps_1$ inside the logarithm is negligible for large $n$, we finally obtain the result of \eqref{eq:dec-simple-asymptotics}.
\end{IEEEproof}

Note that in the discussion above, we did not make any assumptions on $p$. In fact, both Theorems~\ref{thm:dec-simple-informed} and \ref{thm:dec-simple-asymptotics} hold for \textit{arbitrary} values of $p$; the decoder always achieves the capacity associated to that value of $p$. As a result, if we optimize and fix $p$ based on $\vec{\theta}$ (using results from~\cite[Section~II]{laarhoven14capacities}), we automatically end up with a decoder that provably achieves capacity for this attack.

%WWWWWWWWWWWWWWWWWWWWWWWWWWWWWWWWWWWWWWWWWWWWWWWWWWWWWWWW

\subsection{Fingerprinting attacks}
\label{sec:dec-simple-fp}

Let us now consider specific pirate attacks which are often considered in the fingerprinting literature, and investigate the resulting code lengths. We will again consider the following five attacks of \cite{laarhoven14capacities}, which were also considered in e.g. \cite{charpentier09, furon09b, huang12, laarhoven13wifs, meerwald11b, meerwald12, oosterwijk13b}. Recall that $\theta_z = f_{Y|Z}(1|z)$ is the probability of the pirates outputting a $1$ when they received $z$ ones.
\begin{itemize}
  \item \textbf{Interleaving attack}: The coalition randomly selects a pirate and outputs his symbol. This corresponds to 
\begin{align}
(\thint)_z = \frac{z}{c} \, . \qquad (0 \leq z \leq c)
\end{align}
  %This attack is known to be asymptotically optimal in the uninformed max-min fingerprinting game~\cite{huang12}.
  \item \textbf{All-$1$ attack}: The pirates output a $1$ whenever they can, i.e. whenever they have at least one $1$. This translates to 
\begin{align}
(\thall)_z = \begin{cases} 0 & \text{if } z = 0; \\ 1 & \text{if } z > 0. \end{cases}
\end{align}
  \item \textbf{Majority voting}: The colluders output the most common received symbol. This corresponds to
\begin{align}
(\thmaj)_z = \begin{cases} 0 & \text{if } z < \frac{c}{2}; \\ 1 & \text{if } z > \frac{c}{2}. \end{cases}
\end{align}
  \item \textbf{Minority voting}: The colluders output the least common received symbol. This corresponds to
\begin{align}
(\thmin)_z = \begin{cases} 0 & \text{if } z = 0 \text{ or } \frac{c}{2} < z < c; \\ 1 & \text{if } z = c \text{ or } 0 < z < \frac{c}{2}. \end{cases}
\end{align}
  \item \textbf{Coin-flip attack}: If the pirates receive both symbols, they flip a fair coin to decide which symbol to output:
\begin{align}
(\thcoi)_z = \begin{cases} 0 & \text{if } z = 0; \\ \frac{1}{2} & \text{if } 0 < z < c; \\ 1 & \text{if } z = c. \end{cases}
\end{align}
\end{itemize}
Using Theorem~\ref{thm:dec-simple-informed} we can now obtain exact, provable expressions for the code length $\ell$ in terms of $\vec{\theta}, p, c, n, \eps_1, \eps_2$. In general these expressions are quite ugly, but performing a Taylor series expansion around $c = \infty$ for the optimal values of $p$ from \cite[Section II.A]{laarhoven14capacities} we obtain the following expressions for $\ell$. Note that $\ell(\thmin) \sim \ell(\thall)$.
\begin{align}
\ell(\thint) &= 2 c^2 \ln\left(\frac{n}{\eps_1}\right) \left[\frac{1 + \sqrt{\gamma} - \gamma}{1 - \sqrt{\gamma}} + O\left(\frac{1}{c^2}\right)\right], \label{eq:dec-simple-int} \\
\ell(\thall) &= \frac{c \ln\left(\frac{n}{\eps_1}\right)}{\ln(2)^2}  \left[\frac{\sqrt{\gamma}\left(1 + \sqrt{\gamma} - \gamma\right)\ln 2}{1 - 2^{-\sqrt{\gamma}}} + O\left(\frac{1}{c}\right)\right], \\
\ell(\thmaj) &= \pi c \ln\left(\frac{n}{\eps_1}\right) \left[\frac{1 + \sqrt{\gamma} - \gamma}{1 - \sqrt{\gamma}} + O\left(\frac{1}{c}\right)\right], \\
\ell(\thcoi) &= \frac{4c}{\ln(2)^2} \ln\left(\frac{n}{\eps_1}\right) \\
 &\times \left[\frac{\sqrt{\gamma} \left(1 + \sqrt{\gamma} - \gamma\right) \ln 2 }{2 - \left(1+\frac{1}{\sqrt{2}}\right)^{\sqrt{\gamma}} - \left(1-\frac{1}{\sqrt{2}}\right)^{\sqrt{\gamma}}} + O\left(\frac{1}{c}\right)\right].
\end{align}
If we assume that both $c \to \infty$ and $\gamma \to 0$, then we can further simplify the above expressions for the code lengths. The first terms between brackets all scale as $1 + O(\sqrt{\gamma})$, so the code lengths scale as the terms before the square brackets. These code lengths match the capacities of \cite{laarhoven14capacities}.

As for the score functions $g$, let us highlight one attack in particular, the interleaving attack. The all-$1$ decoder will be discussed in Section~\ref{sec:dec-simple-gt}, while the score functions for other attacks can be computed in a similar fashion. For the interleaving attack, working out the probabilities in \eqref{eq:dec-simple-g}, we obtain the following score function:
\begin{align}
g(x,y,p) = \begin{cases} 
\ln\left(1 + \frac{p}{c(1 - p)}\right) & \text{if } x = y = 0; \\
\ln\left(1 - \frac{1}{c}\right) & \text{if } x \neq y; \\
\ln\left(1 + \frac{1 - p}{c p}\right) & \text{if } x = y = 1. \end{cases}
\end{align}

%WWWWWWWWWWWWWWWWWWWWWWWWWWWWWWWWWWWWWWWWWWWWWWWWWWWWWWWW

\subsection{Group testing models}
\label{sec:dec-simple-gt}

For group testing, we will consider three models: the classical (noiseless) model and the models with additive noise and dilution noise. Other models where the probability of a positive test result only depends on the tally $Z$ (such as the threshold group testing models considered in~\cite{laarhoven14capacities}) may be analyzed in a similar fashion. Note that the classical model is equivalent to the all-$1$ attack in fingerprinting, as was previously noted in e.g.~\cite{laarhoven13allerton, meerwald11b, stinson00}.
\begin{itemize}
  \item \textbf{Classical model}: The test output is positive iff the tested pool contains at least one defective:
\begin{align}
(\thall)_z = \begin{cases} 0 & \text{if } z = 0; \\ 1 & \text{if } z > 0. \end{cases}
\end{align}
  \item \textbf{Additive noise model}: Just like the classical model, but if no defectives are tested the result may still be positive:
\begin{align}
(\thadd)_z = \begin{cases} r & \text{if } z = 0; \\ 1 & \text{if } z > 0. \end{cases} \qquad (r \in (0,1))
\end{align}
  \item \textbf{Dilution noise model}: Similar to the classical model, but the probability of a positive result increases with $z$:
\begin{align}
(\thdil)_z = \begin{cases} 0 & \text{if } z = 0; \\ 1 - r^z & \text{if } z > 0. \end{cases} \qquad (r \in (0,1))
\end{align}
\end{itemize}
We can again expand the expressions of Theorem~\ref{thm:dec-simple-informed} around $c = \infty$ for the optimal values of $p$ from \cite[Section II.B]{laarhoven14capacities}, but with the added parameter $r$ the resulting formulas are quite a mess. If we also let $\gamma \to 0$ then we can use Theorem~\ref{thm:dec-simple-asymptotics} to obtain the following simpler expressions:
\begin{align}
\ell(\thall) &= \frac{c \ln n}{\ln(2)^2} \left[1 + O\left(\sqrt{\gamma} + \frac{1}{c}\right)\right], \\
\ell(\thadd) &= \frac{c \ln n}{\ln(2)^2 - r \ln 2 + O(r^2)} \left[1 + O\left(\sqrt{\gamma} + \frac{1}{c}\right)\right], \\
\ell(\thdil) &= \frac{c \ln n}{\ln(2)^2 - O(r \ln r)} \left[1 + O\left(\sqrt{\gamma} + \frac{1}{c}\right)\right].
\end{align}
For more detailed expressions for $\ell$, one may combine Theorems~\ref{thm:dec-simple-informed} and \ref{thm:dec-simple-asymptotics} with~\cite[Section II.B]{laarhoven14capacities}. For the classical model, working out the details we obtain the following result.

\begin{corollary} \label{thm:dec-simple-cla}
For the classical group testing model, the simple decoder for the optimal value $p = \palls \approx \frac{\ln 2}{c}$ is given by\footnote{To be precise, for convenience we have scaled $g$ by a factor $(c \ln 2)$, and so also $\eta$ should be scaled by a factor $(c \ln 2)$.}
\begin{align}
g(x,y,\palls) = \begin{cases}
+1 & \text{if } (x,y) = (0,0); \\
-1 + O\left(\frac{1}{c}\right) & \text{if } (x,y) = (0,1); \\
-\infty & \text{if } (x,y) = (1,0); \\
+c & \text{if } (x,y) = (1,1). \end{cases}
\end{align}
Using this decoder in combination with the parameters $\eta$ and $\ell$ of Theorem~\ref{thm:dec-simple-informed}, we obtain a simple group testing algorithm with an optimal asymptotic number of group tests of 
\begin{align}
\ell \sim \frac{c \ln n}{\ln(2)^2} \approx 2.08 c \ln n \approx 1.44 c \log_2 n.
\end{align}
\end{corollary}

This asymptotically improves upon results of e.g.\ Chan et al.~\cite{chan11, chan12} who proposed an algorithm with an asymptotic code length of $\ell \sim e c \ln n \approx 2.72 c \ln n$. Their algorithm does have a guarantee of never falsely identifying a non-defective item as defective (whereas our proposed decoder does not have this guarantee), but the price they pay is a higher asymptotic number of tests to find the defectives.

%WWWWWWWWWWWWWWWWWWWWWWWWWWWWWWWWWWWWWWWWWWWWWWWWWWWWWWWW
%WWWWWWWWWWWWWWWWWWWWWWWWWWWWWWWWWWWWWWWWWWWWWWWWWWWWWWWW

\section{Simple universal decoding}
\label{sec:dec-simple-universal}

While in the previous section we discussed simple decoders for the setting where $\vec{\theta}$ is completely known to the decoder, let us now consider the setting which is more common in fingerprinting, where the attack strategy is not assumed known to the decoder. This setting may partially apply to group testing as well (where there may be some unpredictable noise on the test outputs) but the main focus of this section is the uninformed fingerprinting game.

%WWWWWWWWWWWWWWWWWWWWWWWWWWWWWWWWWWWWWWWWWWWWWWWWWWWWWWWW

\subsection{The simple interleaving decoder, revisited}
\label{sec:dec-simple-universal-intro}

Let us now try to investigate how to build a decoder that works against arbitrary attacks in fingerprinting. To build such a decoder, Meerwald and Furon~\cite{meerwald12} previously noted that Abbe and Zheng~\cite{abbe10} proved in a more general context that under certain conditions on the set of allowed pirate strategies $\vec{\Theta}$, a decoder that works against the worst-case attack $\vec{\theta}^* \in \vec{\Theta}$ also works against arbitrary other attacks $\vec{\theta} \in \vec{\Theta}$. In this case the set of allowed pirate strategies we consider is the set of all attacks satisfying the marking assumption $\pmark$:
\begin{align}
\pmark = \{\vec{\theta} \in [0,1]^{c + 1} \mid \theta_0 = 0, \theta_c = 1\}.
\end{align} 

For finite $c$ the worst-case attack in fingerprinting (from an information-theoretic perspective) has been studied in e.g.~\cite{huang12, meerwald12}, but in general this attack is quite messy and unstructured. Since this attack is not so easy to analyze, let us therefore focus on the asymptotics of large $c$ and $n$. Huang and Moulin~\cite{huang12} previously proved that for large coalitions, the optimal pirate attack is the interleaving attack. So combining this knowledge with the result of Abbe and Zheng, perhaps a good choice for a universal decoder is the interleaving decoder, which we recall is given by:
\begin{align}
g(x,y,p) = \begin{cases} 
\ln\left(1 + \frac{p}{c(1 - p)}\right) & \text{if } x = y = 0; \\
\ln\left(1 - \frac{1}{c}\right) & \text{if } x \neq y; \\
\ln\left(1 + \frac{1 - p}{c p}\right) & \text{if } x = y = 1. \end{cases} \label{eq:dec-simple-universal}
\end{align}
Let us take a closer look at this decoder. For fixed $\delta > 0$, if we look at values $p \in [\delta, 1 - \delta]$ and focus on the regime of large $c$, we can perform a Taylor series expansion around $c = \infty$ to get $\ln(1 + x) \sim x$. The resulting expressions then turn out to be closely related to Oosterwijk et al.'s~\cite{oosterwijk13b} decoder $h$:
\begin{align}
g(x,y,p) \sim \frac{h(x,y,p)}{c} = \begin{cases} 
+\frac{p}{c(1 - p)} & \text{if } x = y = 0; \\
-\frac{1}{c} & \text{if } x \neq y; \\
+\frac{1 - p}{cp} & \text{if } x = y = 1. \end{cases}
\end{align}
This implies that $g$ and $h$ are asymptotically equivalent for $p$ sufficiently far away from $0$ and $1$. Since for Oosterwijk et al.'s score function one generally uses \textit{cut-offs} on $f_P$ (i.e.\ only using values $p \in [\delta, 1 - \delta]$ for fixed $\delta > 0$) to guarantee that $h(x,y,p) = o(c)$ (cf.~\cite{ibrahimi13}), and since the decoder of Oosterwijk et al.\ is known to achieve capacity using these cut-offs, we immediately get the following result.

\begin{proposition} \label{prop:dec-simple-uninformed}
The score function $g$ of \eqref{eq:dec-simple-universal} together with the bias density function (encoder) $f_P^{(\delta)}$ on $[\delta, 1 - \delta]$ of the form
\begin{align}
f_P^{(\delta)}(p) = \frac{1}{(\pi - 4\arcsin \sqrt{\delta})\sqrt{p(1 - p)}} \label{eq:arcsine-cutoffs}
\end{align}
asymptotically achieve the simple capacity for the uninformed fingerprinting game when the same cut-offs $\delta$ as those of~\cite{ibrahimi13} are used.
\end{proposition}

So combining the log-likelihood decoder tuned against the asymptotic worst-case attack (the interleaving attack) with the arcsine distribution with cut-offs, we obtain a universal decoder that works against arbitrary attacks.

%WWWWWWWWWWWWWWWWWWWWWWWWWWWWWWWWWWWWWWWWWWWWWWWWWWWWWWWW

\subsection{Cutting off the cut-offs}
\label{sec:dec-simple-universal-cutoffs}

Although Proposition~\ref{prop:dec-simple-uninformed} is already a nice result, the cut-offs $\delta$ have been a nagging inconvenience ever since Tardos introduced them in 2003~\cite{tardos03}. In previous settings it was well-known that this cut-off $\delta$ had to be large enough to guarantee that innocent users are not falsely accused, and small enough to guarantee that large coalitions can still be caught. For instance, when using Tardos' original score function, it was impossible to do without cut-offs, and the same seems to hold for Oosterwijk et al.'s decoder $h$, since the scores blow up for $p \approx 0,1$.

Looking at the universal log-likelihood decoder, one thing to notice is that the logarithm has a kind of mitigating effect on the tails of the score distributions. For $0 \ll p \ll 1$ the resulting scores are roughly a factor $c$ smaller than those obtained with $h$, but where the blow-up effect of $h$ for small $p$ is proportional to $\frac{1}{p}$, the function $g$ only scales as $\ln(\frac{1}{p})$ in the region of small $p$. This motivates the following claim, showing that with this decoder $g$ we finally do not need any cut-offs anymore! 

\begin{theorem} \label{thm:dec-simple-universal}
The decoder $g$ of \eqref{eq:dec-simple-universal} and the encoder $f_P^*(p)$, defined on $[0,1]$ by
\begin{align}
f_P^*(p) = \frac{1}{\pi\sqrt{p(1 - p)}}, \quad \label{eq:arcsine}
\end{align}
together asymptotically achieve the simple capacity for the uninformed fingerprinting game.
\end{theorem}

\begin{IEEEproof}
We will argue that using this new universal decoder $g$, the difference in performance between using and not using cut-offs on $f_P$ is negligible for large $c$. Since the encoder with cut-offs asymptotically achieves capacity, it then follows that without cut-offs this scheme also achieves capacity. 

Let us first prove that all moments of user scores are finite, even if no cut-offs are used. We will show that $\expn[g(x,y,p)] < \infty$ for any $x$ and $y$, so that after taking weighted combinations we also get $\expn(S_{j,i} | H_{0/1}) < \infty$. Let us consider the case where $x = y = 1$; other cases can be analyzed in the same way. Using the density function $f_P^*$ of~\eqref{eq:arcsine}, we have
\begin{align}
E &= \expn[g(1,1,p)^k] \\
 &= \int_{0}^1 \frac{dp}{\pi \sqrt{p(1-p)}} \log^k\left(1 + \frac{1 - p}{c p}\right).
\end{align}
Splitting the interval $[0,1]$ into two parts $[\delta, 1]$ and $[0,\delta]$ (where $\delta$ depends on $k$ but not on $c$) we obtain
\begin{align}
E &= \int_{\delta}^1 \frac{dp}{\pi \sqrt{p(1-p)}} \log^k\left(1 + \frac{1 - p}{c p}\right) \\
 &+ \int_{0}^{\delta} \frac{dp}{\pi \sqrt{p(1-p)}} \log^k\left(1 + \frac{1 - p}{c p}\right).
\end{align}
Let us denote the two terms by $E_1$ and $E_2$. For the first term, we can perform a Taylor series expansion to obtain:
\begin{align}
E_1 &= \int_{\delta}^1 \frac{dp}{\pi \sqrt{p(1-p)}} \log^k\left(1 + \frac{1 - p}{c p}\right) \\
 &= \int_{\delta}^1 \frac{dp}{\pi \sqrt{p(1-p)}} \left(\frac{1 - p}{c p} + O\left(\frac{(1 - p)^2}{c^2 p^2}\right)\right)^k \\
 &\leq \int_{\delta}^1 \frac{dp}{\pi \sqrt{p(1-p)}} \left(\frac{1}{c \delta} + O\left(\frac{1}{c^2 \delta^2}\right)\right)^k \\
 &\stackrel{(a)}{\leq} \int_{\delta}^1 \frac{dp}{\pi \sqrt{p(1-p)}} = 1 < \infty.
\end{align}
Here $(a)$ follows from considering sufficiently large $c$ while $\delta$ remains fixed. (Note that for large $c$ we even have $E_1 \to 0$.) For the other term we do not expand the logarithm:
\begin{align}
E_2 &= \int_0^{\delta} \frac{dp}{\pi \sqrt{p(1-p)}} \log^k\left(1 + \frac{1 - p}{c p}\right) \\
 &\propto \int_0^{\delta} \frac{dp}{\sqrt{p}} \log^k\left(\frac{1}{p}\right) \stackrel{(b)}{\to} 0. \label{eq:convergence}
\end{align}
The last step $(b)$ follows from the fact that the integration is done over an interval of width $\delta$, while the integrand scales as $\frac{1}{\sqrt{p}}$ times some less important logarithmic terms. For arbitrary $k$, we can thus let $\delta = \delta(k) \to 0$ as a function of $k$ to see that this is always bounded. Similar arguments can be used to show that for other values of $x,y$ we also have $\expn[g(x,y,p)^k] < 0$.

As a result, all innocent and guilty user score moments are finite, and so for large $c$ from the Central Limit Theorem it follows that the distributions of user scores will converge to Gaussians. If the scores of innocent and guilty users are indeed Gaussian for large $c$, then as discussed in e.g.~\cite{oosterwijk13b, skoric08} all that matters for assessing the performance of the scheme are the mean and variance of both curves. Similar to \eqref{eq:convergence}, the effects of small cut-offs on the distribution function $f_P$ are negligible as both means and variances stay the same up to small order terms. So indeed, in both cases the `performance indicator'~\cite{oosterwijk13b} asymptotically stays the same, leading to equivalent code lengths.
\end{IEEEproof}

\begin{figure*}[!ht]
\centering
\subfloat[][The PDF of innocent user scores using the score function $g$]{\includegraphics[width=0.45\textwidth]{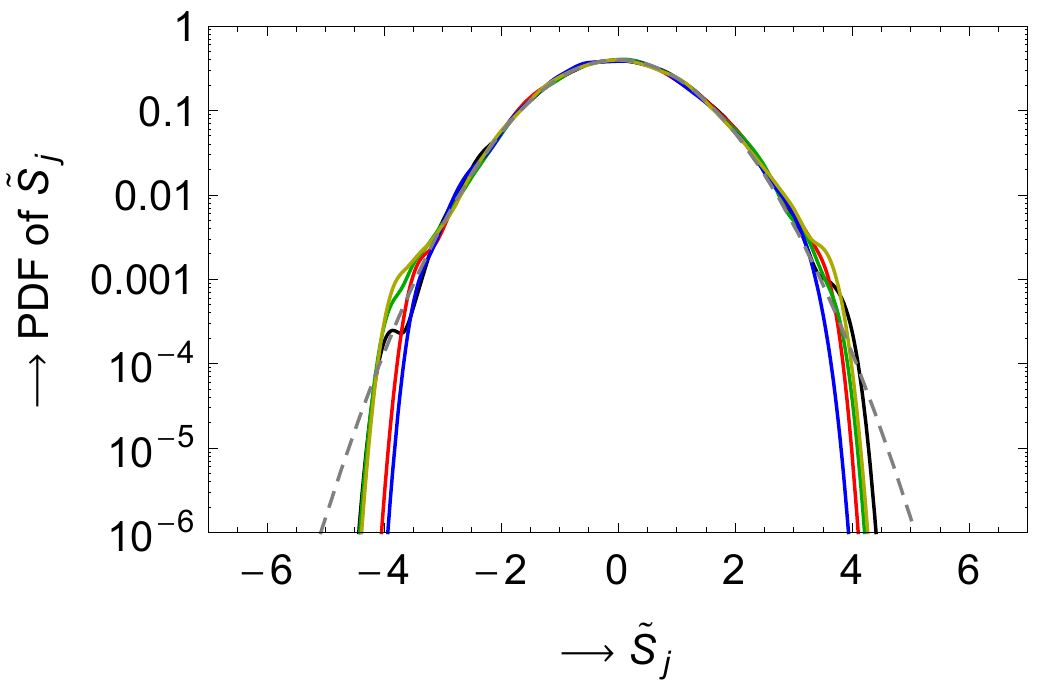}\label{fig:pdf-g}} \,
\subfloat[][The PDF of innocent user scores using the score function $h$]{\includegraphics[width=0.45\textwidth]{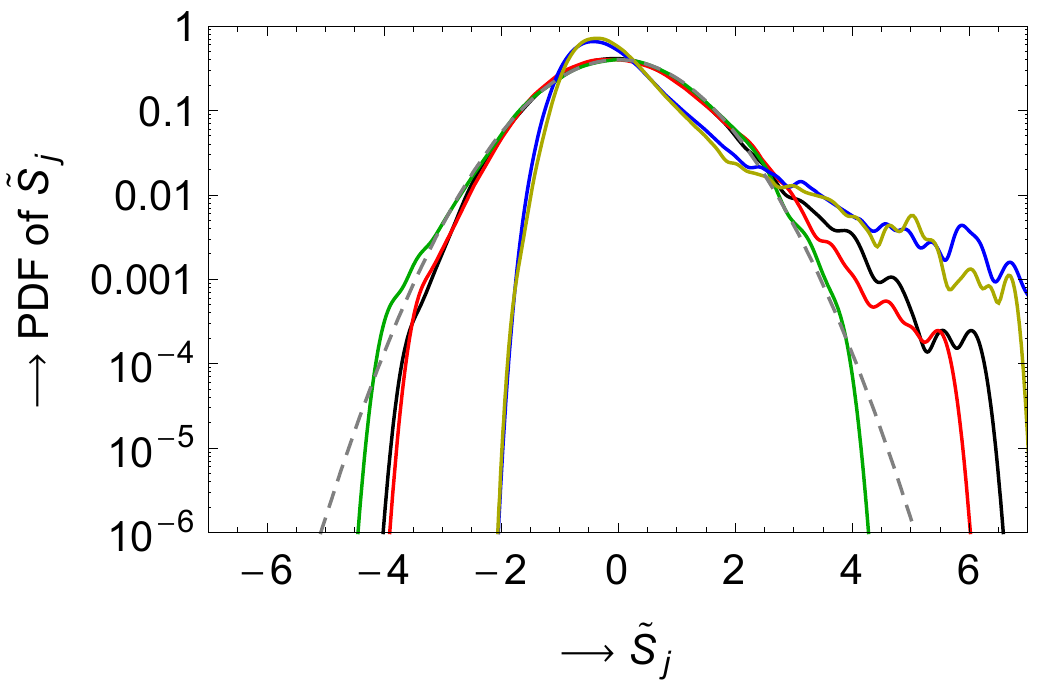}\label{fig:pdf-h}}
\caption{Estimates of the probability density functions of normalized innocent user scores for $c = 10$ and $\ell = 10\,000$ based on $10\,000$ simulations of innocent user scores for each attack. The different lines correspond to the interleaving attack (black), all-$1$ attack (red), majority voting (green), minority voting (blue), and the coin-flip attack (yellow), while the gray dashed line corresponds to the standard normal distribution with $f_Z(z) \propto \exp(-\tfrac{1}{2}z^2)$. Note that the drops near densities of $10^{-4}$ (the far ends of the curves) occur due to smoothening the curve with only $10^4$ samples, and do not reflect the real distribution tails. \label{fig:pdf}}
\end{figure*}

Note that the same result does \textit{not} apply to the score function $h$ of Oosterwijk et al.~\cite{oosterwijk13b}, for which the effects of values $p \approx 0,1$ are not negligible. The main difference is that for small $p$, the score function $h$ scales as $\frac{1}{p}$ (which explodes when $p$ is really small), while the log-likelihood decoder $g$ only scales as $\ln(\tfrac{1}{p})$. Figure~\ref{fig:pdf} illustrates the difference in the convergence of normalized innocent user scores to the standard normal distribution, when using the score functions $g$ and $h$. These are experimental results for $c = 10$ and $\ell = 10\,000$ based on $10\,000$ simulated scores for each curve, and for both score functions we did not use any cut-offs. As we can see, using $g$ the normalized scores $\tilde{S}_j = (S_j - \expn S_j) / \sqrt{\var S_j}$ are close to Gaussian, while using $h$ the curves especially do not look very Gaussian for $\tilde{S}_j \gg 0$; in most cases the distribution tails are much too large. For the minority voting attack the resulting curve does not even seem close to a standard normal Gaussian distribution.

%WWWWWWWWWWWWWWWWWWWWWWWWWWWWWWWWWWWWWWWWWWWWWWWWWWWWWWWW

\subsection{Designing the scheme}
\label{sec:dec-simple-universal-design}

With the above result in mind, let us now briefly discuss how to actually build a universal scheme with the interleaving decoder $g$. From Theorem~\ref{thm:dec-simple-universal} it is clear that for generating biases we should use the arcsine distribution $f_P^*$, and our decoder will be the interleaving decoder $g$ of~\eqref{eq:dec-simple-universal}. What remains is figuring out how to choose $\ell$ and $\eta$ for arbitrary attacks.

First, it is important to note that the expected innocent and guilty scores per segment ($\mu_1 = \expn(S_{j,i}|H_1)$ and $\mu_0 = \expn(S_{j,i}|H_0)$) and the variance of the innocent and guilty scores ($\sigma^2_1 = \var(S_{j,i}|H_1)$ and $\sigma^2_0 = \var(S_{j,i}|H_0)$) heavily depend on the collusion channel $\vec{\theta}$. This was not the case for Tardos' original decoder~\cite{tardos03} and the symmetrized decoder~\cite{skoric08}, for which $\eta$ could be fixed in advance regardless of the collusion strategy. This means that we will either have to delay fixing $\eta$ until the decoding stage, or scale/translate scores per segment accordingly at each position $i$.

For choosing the code length $\ell$ and threshold $\eta$ let us focus on the regime of reasonably large $c$. In that case, as argued above the total innocent and guilty scores will behave like Gaussians, with parameters $S_1 \sim \mathcal{N}(\ell \mu_1, \ell \sigma_1^2)$ and $S_0 \sim \mathcal{N}(\ell \mu_0, \ell \sigma_0^2)$. To distinguish between these two distributions, using e.g.\ Sanov's theorem the code rate $\ell/\ln n$ should be proportional to the Kullbeck-Leibler divergence between the two distributions:
\begin{align}
d(S_0 \| S_1) = \frac{(\mu_0 - \mu_1)^2}{\sigma_1^2} + \frac{1}{2}\left(\frac{\sigma_0^2}{\sigma_1^2} - 1 - \ln \frac{\sigma_0^2}{\sigma_1^2}\right).
\end{align}
A similar expression appears in \cite{skoric08, oosterwijk13b}, where it was noted that $\sigma_0 \ll \sigma_1$, so that the first term is the most important term. In \cite{ibrahimi13, oosterwijk13b} the ratio $\frac{(\mu_0 - \mu_1)^2}{\sigma_1^2}$ was coined the `performance indicator', and it was argued that this ratio should be maximized. In \cite{oosterwijk13b} it was further shown that when using their decoder $h$, this ratio is minimized by the pirates when they choose the interleaving attack $\thint$. In other words: assuming scores are Gaussian for large $c$, the best attack the pirates can use is the interleaving attack when using $h$ as the decoder.

Since our new decoder $g$ is very similar to Oosterwijk et al.'s decoder $h$ (by $c \cdot g \approx h$), a natural conjecture would be that also for this new score function, asymptotically the best pirate attack maximizing the decoder's error probabilities is the interleaving attack. Experiments with $g$ and previous experiments of \cite{oosterwijk13b} with $h$ indeed show that other pirate attacks (such as those considered in Section~\ref{sec:dec-simple-fp}) generally perform worse than the interleaving attack. As a result, a natural choice for selecting $\ell$ would be to base $\ell$ on the code length needed to deal with the (asymptotic) worst-case attack for this decoder, which we conjecture is the interleaving attack. And for the interleaving attack we know how to choose $\ell$ by Theorem~\ref{thm:dec-simple-informed}, Equation~\ref{eq:dec-simple-int} and Theorem~\ref{thm:dec-simple-asymptotics}:
\begin{align}
\ell &= \frac{\sqrt{\gamma} (1 + \sqrt{\gamma} - \gamma)}{-\ln M(1 - \sqrt{\gamma})} \, \ln\left(\frac{n}{\eps_1}\right) \label{eq:ell-int-uni} \\
 &= 2 c^2 \ln\left(\frac{n}{\eps_1}\right) \left[\frac{1 + \sqrt{\gamma} - \gamma}{1 - \sqrt{\gamma}} + O\left(\frac{1}{c^2}\right)\right] \label{eq:ell-int-uni2} \\
 &= 2 c^2 \ln n \left[1 + O\left(\sqrt{\gamma} + \frac{1}{c^2}\right)\right]. \label{eq:ell-int-uni3}
\end{align}
These choices for $\ell$ thus seem reasonable estimates for the code lengths required to deal with arbitrary attacks.

Finally, for choosing $\eta$, as argued before this parameter depends on the pirate strategy $\vec{\theta}$, which may lead to different scalings and translations of the curves of innocent and guilty user scores. What we could do is compute the parameters $\mu_{0/1}, \sigma^2_{0/1}$ based on the pirate output $\vec{y}$ and normalize the scores accordingly. This means that after computing user scores $S_j$, we apply the following transformation:
\begin{align}
\tilde{S}_j &= \frac{S_j - \ell \mu_1}{\sqrt{\ell} \sigma_1}.
\end{align}
This guarantees that the scores of innocent users will roughly be distributed as $\mathcal{N}(0,1)$, and for guilty users this results in a distribution of the form $\mathcal{N}(\sqrt{\ell}(\frac{\mu_0 - \mu_1}{\sigma_1}),\frac{\sigma_0^2}{\sigma_1^2})$. To guarantee that the probability that an innocent user has a score below $\eta$ is at least $1 - \frac{\eps_1}{n}$, it then suffices to let
\begin{align}
\tilde{\eta} = \frac{\eta - \ell \mu_1}{\sqrt{\ell} \sigma_1} \approx \Phi^{-1}\left(1 - \frac{\eps_1}{n}\right) \sim \sqrt{\ln\left(\frac{n}{\eps_1}\right)},
\end{align} 
where $\Phi$ denotes the distribution function of the standard normal distribution $\mathcal{N}(0,1)$. This means that after transforming the scores, the threshold can be fixed independent of the pirate strategy.

%As an example, consider a practical scenario with $n = 10^6$ users and $c = 25$ colluders, where we want to put bounds on the error probabilities as $\eps_1 = \eps_2 = 10^{-3}$. Using the leading term for the code length for the interleaving attack of \eqref{eq:ell-int-uni2}, we get $\ell \approx 76\,000$ when we set $\gamma = \ln(1 / \eps_2)/\ln(n / \eps_1)$ (as for the provable bounds of Theorem~\ref{thm:dec-simple-informed} for catching at least one colluder), and we get $\ell \approx 104\,000$ when we set $\gamma = \ln(c / \eps_2)/\ln(n / \eps_1)$ (as discussed in Section~\ref{sec:dec-simple-practice}, for catching all colluders). 

%WWWWWWWWWWWWWWWWWWWWWWWWWWWWWWWWWWWWWWWWWWWWWWWWWWWWWWWW

\subsection{Another simple universal decoder}
\label{sec:dec-simple-universal-emi}

Besides Oosterwijk et al.'s Lagrangian approach and our Neyman-Pearson-based approach to obtaining efficient decoders, let us now mention a third way to obtain a similar capacity-achieving universal decoder. 

To construct this decoder, we start with the empirical mutual information decoder proposed by Moulin~\cite{moulin08}, and for now let us assume $p_i \equiv p$ is fixed\footnote{When $p_i$ is not fixed and is drawn from a continuous distribution function $f_P$, the empirical probabilities considered below do not make much sense, as each value of $p_i$ only occurs once. In that case one could e.g.\ build a histogram for the values of $p$, and compute empirical probabilities for each bin, or discretize the distribution function for $p$~\cite{laarhoven13ihmmsec, nuida09}.}. With this decoder, a user is assigned a score of the form
\begin{align}
S_j &= \sum_{x,y} \hat{f}_{X,Y|P}(x,y|p) \ln\left(\frac{\hat{f}_{X,Y|P}(x,y|p)}{\hat{f}_{X|P}(x|p)\hat{f}_{Y|P}(y|p)}\right)
\end{align}
and again the decision to accuse depends on whether this score exceeds some fixed threshold $\eta$. Here $\hat{f}$ is the empirical estimate of the actual probability $f$, i.e., $\hat{f}_{X,Y|P}(x,y|p) = |\{i: (x_{j,i}, y_i) = (x,y)\}| / \ell$. Writing out the empirical probability outside the logarithm, and replacing the summation over $x,y$ by a summation over the positions $i$, this is equivalent to
\begin{align}
S_j %&= \frac{1}{\ell} \sum_{i=1}^{\ell} \ln\left(\frac{\hat{f}_{X,Y|P}(x_{j,i},y_i|p)}{\hat{f}_{X|P}(x_{j,i}|p)\hat{f}_{Y|P}(y_i|p)}\right) \\
 &= \frac{1}{\ell} \sum_{i=1}^{\ell} \ln\left(\frac{\hat{f}_{X,Y|P}(x_{j,i},y_i|p)}{\hat{f}_{X,Y|P}(x_{j,i},y_i|p, H_1)}\right).
\end{align}
Now, this almost fits the score-based simple decoder framework, except for that the terms inside the logarithm are not independent for different positions $i$. To overcome this problem, we could try to replace the empirical probabilities $\hat{f}$ by the actual probabilities $f$, but to compute $f_{X,Y|P}(x_{j,i}, y_i|p)$ we need to know whether user $j$ is guilty or not. Solving this final problem using Bayesian inference, we get the following result.

\begin{lemma}
Approximating the empirical probabilities in the empirical mutual information decoder with actual probabilities using Bayesian inference (with an a priori probability of guilt of $\pr(j \in \mathcal{C}) = \frac{c}{n}$), this decoder corresponds to using the following score function $m$:
\begin{align}
m(x,y,p) = \ln\left(1 + \frac{c}{n}\left[\frac{f_{X,Y|P}(x,y|p,H_0)}{f_{X,Y|P}(x,y|p,H_1)} - 1\right]\right). \label{eq:dec-simple-moulin1}
\end{align}
\end{lemma}

\begin{IEEEproof}
The value of $f_{X,Y|P}(x_{j,i},y_i|p,H_1)$ can be computed without any problems, so let us focus on the term $f_{X,Y|P}(x_{j,i},y_i|p)$. Using Bayesian inference, we have:
\begin{align}
f_{X,Y|P}(x,y|p) &= \pr(j \in \mathcal{C}) f_{X,Y|P}(x,y|p,H_0) \\
&+ \pr(j \notin \mathcal{C}) f_{X,Y|P}(x,y|p,H_1).
\end{align}
Assuming an a priori probability of guilt of $\pr(j \in \mathcal{C}) = \frac{c}{n}$ and dividing by $f_{X,Y|P}(x,y|p,H_1)$ we get
\begin{align}
\frac{f_{X,Y|P}(x,y|p)}{f_{X,Y|P}(x,y|p,H_1)} &= \frac{c}{n} \frac{f_{X,Y|P}(x,y|p,H_0)}{f_{X,Y|P}(x,y|p,H_1)}  + 1 - \frac{c}{n}.
\end{align}
Taking logarithms, this leads to the expression of \eqref{eq:dec-simple-moulin1}.
\end{IEEEproof}

Although this score function looks very similar to the log-likelihood decoder, there are some essential differences. For instance, for the all-$1$ attack we have $g(1,0,p) = -\infty$ while $m(1,0,p) = \ln(1 - \frac{c}{n}) > -\infty$. For the interleaving attack, for which we may again hope to obtain a universal decoder using this approach, we do get a familiar result.

\begin{corollary} \label{thm:dec-simple-moulin}
For the interleaving attack, the Bayesian approximation of the empirical mutual information decoder of \eqref{eq:dec-simple-moulin1} satisfies
\begin{align}
m(x,y,p) = \begin{cases}
\ln\left(1 + \frac{p}{n(1 - p)}\right) & \text{if } x = y = 0; \\
\ln\left(1 - \frac{1}{n}\right) & \text{if } x \neq y; \\ 
\ln\left(1 + \frac{1 - p}{np}\right) & \text{if } x = y = 1.
\end{cases} \label{eq:dec-simple-moulin2}
\end{align}
\end{corollary}

For values of $p \in [\delta, 1 - \delta]$ with $\delta > 0$ and large $c$, this decoder is again equivalent to both the log-likelihood score function $g$ and Oosterwijk et al.'s score function $h$:
\begin{align}
c \cdot g(x,y,p) \sim n \cdot m(x,y,p) \sim h(x,y,p).
\end{align}
For $p \approx 0,1$ the logarithm again guarantees that scores do not blow up so much, but due to the factor $n$ in the denominator (rather than a factor $c$, as in $g$) the scores relatively increase more when $p$ approaches $0$ than for the score function $g$.

%WWWWWWWWWWWWWWWWWWWWWWWWWWWWWWWWWWWWWWWWWWWWWWWWWWWWWWWW
%WWWWWWWWWWWWWWWWWWWWWWWWWWWWWWWWWWWWWWWWWWWWWWWWWWWWWWWW
%WWWWWWWWWWWWWWWWWWWWWWWWWWWWWWWWWWWWWWWWWWWWWWWWWWWWWWWW
%WWWWWWWWWWWWWWWWWWWWWWWWWWWWWWWWWWWWWWWWWWWWWWWWWWWWWWWW

\section{Joint informed decoding}
\label{sec:dec-joint}

In this section we will discuss informed joint decoders which we conjecture are able to find pirates with shorter code lengths than simple decoders. The asymptotics of the resulting code lengths further motivate that these schemes may be optimal, but some open problems remain for proving that they are indeed optimal.

Following the score-based framework for joint decoders of Moulin~\cite{moulin08}, we assign tuples $T$ of size $c$ a score of the form
\begin{align}
S_T = \sum_{i=1}^{\ell} S_{T,i} = \sum_{i=1}^{\ell} g(\{x_{j,i}: j \in T\}, y_i, p_i).
\end{align}
Note that if $\vec{\theta}$ is colluder-symmetric, then this is equivalent to
\begin{align}
S_T = \sum_{i=1}^{\ell} g(z_{T,i}, y_i, p_i),
\end{align}
where $z_{T,i} = \sum_{j \in T} x_{j,i}$ is the tally of the number of ones received by the tuple $T$ in position $i$. For the accusation phase, we now accuse all users in $T$ iff $S_T \geq \eta$ for some fixed threshold $\eta$. Note that this accusation algorithm is not exactly well-defined, since it is possible that a user appears both in a tuple that is accused and in a tuple that is not accused. For the analysis we will assume that the scheme is only successful if the single tuple consisting of all colluders has a score exceeding $\eta$ and no other tuples have a score exceeding $\eta$, in which case all users in that guilty tuple are accused. This may be too pessimistic for evaluating the performance of the scheme.

%WWWWWWWWWWWWWWWWWWWWWWWWWWWWWWWWWWWWWWWWWWWWWWWWWWWWWWWW

\subsection{Joint log-likelihood decoders}
\label{sec:dec-joint-intro}

For building a joint decoder we would like to be able to distinguish between the all-guilty tuple and other tuples, so a natural generalization of the hypotheses $H_0$ and $H_1$ for simple decoding would be to let $H_0: T = \mathcal{C}$ and $H_1: T \neq \mathcal{C}$. However, with this choice of $H_1$, computing probabilities $f_{Z,Y|P}(z,y|p,H_1)$ is complicated: the event $H_1$ does not completely determine $f_{Y|Z}(y|z)$, since that depends on exactly how many colluders are present in $T$. To be able to compute the likelihood ratios, we therefore use the following two hypotheses, which were also used in e.g.~\cite{meerwald12}:
\begin{align}
&H_0: \quad \text{all users $j \in T$ are guilty } (T = \mathcal{C}), \\
&H_1: \quad \text{all users $j \in T$ are innocent } (T \cap \mathcal{C} = \varnothing).
\end{align}
We again use the corresponding log-likelihood ratio per position as our score function $g$, which is again the logarithm of the likelihood ratio over all positions $i$:
\begin{align}
g(z,y,p) = \ln\left(\frac{f_{Z,Y|P}(z,y|p,H_0)}{f_{Z,Y|P}(z,y|p,H_1)}\right). \label{eq:dec-joint-g}
\end{align}
Using this joint score function $g$ corresponds to a most powerful test according to the Neyman-Pearson lemma~\cite{neyman33}, so $g$ is in a sense optimal for distinguishing between $H_0$ and $H_1$. Joint decoders of this form were previously considered in e.g.~\cite{meerwald12}.

%WWWWWWWWWWWWWWWWWWWWWWWWWWWWWWWWWWWWWWWWWWWWWWWWWWWWWWWW

\subsection{Theoretical evaluation}
\label{sec:dec-joint-theory}

Let us again study how to choose $\ell$ and $\eta$ such that we can prove that the false positive and false negative error probabilities are bounded from above by certain values $\eps_1$ and $\eps_2$. Below we will again make use of the function $M$ of \eqref{eq:M} where the simple hypotheses have been replaced by our new joint hypotheses $H_0$ and $H_1$.

\begin{theorem} \label{thm:dec-joint-informed}
Let $p$ and $\vec{\theta}$ be fixed and known to the decoder. Let $\gamma = \ln(1/\eps_2) / \ln(n^c/\eps_1)$, and let the code length $\ell$ and the threshold $\eta$ be defined as
\begin{align}
\ell &= \frac{\sqrt{\gamma} (1 + \sqrt{\gamma} - \gamma)}{-\ln M(1 - \sqrt{\gamma})} \ln\left(\frac{n^c}{\eps_1}\right). \\
\eta &= (1 - \gamma)\ln\left(\frac{n^c}{\eps_1}\right), \qquad 
\end{align}
Then with probability at least $1 - \eps_1$ all all-innocent tuples are not accused, and with probability at least $1 - \eps_2$ the single all-guilty tuple is accused.
\end{theorem}

\begin{IEEEproof}
The proof is very similar to the proof of Theorem~\ref{thm:dec-simple-informed}. Instead of $n$ innocent and $c$ guilty users we now have $\binom{n}{c} < n^c$ all-innocent tuples and just $1$ all-guilty tuple, which changes some of the numbers in $\gamma$, $\eta$ and $\ell$ above. We then again apply the Markov inequality with $\alpha = 1 - \sqrt{\gamma}$ and $\beta = \sqrt{\gamma}$ to obtain the given expressions for $\eta$ and $\ell$.
\end{IEEEproof}

For deterministic strategies $\vec{\theta} \in \{0,1\}^{c+1}$, choosing the scheme parameters is much simpler. Similar to \cite[Lemma 1]{laarhoven14capacities}, where it was shown that for deterministic attacks the capacity is exactly $\frac{1}{c}$, in this case we get a code length of roughly $c \log_2 n$.

\begin{theorem} \label{thm:dec-joint-deterministic}
Let $\vec{\theta}$ be a deterministic attack, and let $p$ be chosen such that $f_{Y|P}(1|p) = \frac{1}{2}$. Let $\ell$ and $\eta$ be chosen as:
\begin{align}
\ell = \log_2 \left(\frac{n^c}{\eps_1}\right), \quad \eta = \ln \left(\frac{n^c}{\eps_1}\right),
\end{align}
Then with probability $1 - \eps_1$ all all-innocent tuples will not be accused, and the single all-guilty tuple will \textit{always} be accused.
\end{theorem}

\begin{IEEEproof}
For deterministic attacks, we have
\begin{align}
f_{Z,Y|P}(z,y|p,H_0) &= f_{Z|P}(z|p) f_{Y|Z}(y|z), \\
f_{Z,Y|P}(z,y|p,H_1) &= f_{Z|P}(z|p) f_{Y|P}(y|p).
\end{align}
As a result, the score function $g$ satisfies
\begin{align}
g(z,y,p) = \begin{cases}
-\ln f_{Y|P}(y|p) & \text{if } \theta_z = y; \\
-\infty & \text{if } \theta_z \neq y.
\end{cases}
\end{align}
With the capacity-achieving choice of $p$ of \cite[Lemma 1]{laarhoven14capacities}, we have $f_{Y|P}(y|p) = \frac{1}{2}$ for $y = 0,1$ leading to a score of $+\ln 2$ for a match, and $-\infty$ for cases where $y_i$ does not match the output that follows from $\vec{\theta}$ and the assumption that $T$ is the all-guilty tuple. For $T = \mathcal{C}$, clearly we will always have a match, so this tuple's score will always be $\ell \ln 2$, showing that this tuple is always accused. 

For innocent tuples, since $f(z,y|p,H_1) = \frac{1}{2} f(z,y|p,H_0)$ it follows that in each position $i$, with probability $\frac{1}{2}$ this tuple's score will not be $-\infty$. So with probability $2^{-\ell}$, the tuple's score after $\ell$ segments will not be $-\infty$, in which case it equals $\ell \ln 2$. To make sure that this probability is at most $\eps_1/n^c$ so that the total error probability is at most $\eps_1$, we set $2^{-\ell} = \eps_1/n^c$, leading to the resulting expression for $\ell$. 
\end{IEEEproof}

Note that for deterministic attacks, any choice of $-\infty < \eta_0 \leq \eta$ works just as well as choosing $\eta$; after $\ell$ segments all tuples will either have a score of $-\infty$ or $\eta$.

%WWWWWWWWWWWWWWWWWWWWWWWWWWWWWWWWWWWWWWWWWWWWWWWWWWWWWWWW

\subsection{Practical evaluation}
\label{sec:dec-joint-practice}

Theorem~\ref{thm:dec-joint-informed} does not prove that we can actually find the set of colluders with high probability, since mixed tuples (consisting of some innocent and some guilty users) also exist and these may or may not have a score exceeding $\eta$. Theorem~\ref{thm:dec-joint-informed} only proves that with high probability we can find a set $\mathcal{C}'$ of $c$ users which contains at least one colluder. Basic experiments show that in many cases the only tuple with a score exceeding $\eta$ (and thus the tuple with the highest score) is the all-guilty tuple, and so all mixed tuples have a score below $\eta$. Proving that mixed tuples indeed get a score below $\eta$ is left as an open problem.

%WWWWWWWWWWWWWWWWWWWWWWWWWWWWWWWWWWWWWWWWWWWWWWWWWWWWWWWW

\subsection{Asymptotic code lengths}

To further motivate why using this joint decoder may be the right choice, the following proposition shows that at least the resulting code lengths are optimal. The proof is analogous to the proof of Theorem~\ref{thm:dec-simple-asymptotics}.

\begin{theorem} \label{thm:dec-joint-asymptotics}
If $\gamma = o(1)$ then the code length $\ell$ of Theorem~\ref{thm:dec-joint-informed} scales as
\begin{align}
\ell = \frac{\log_2 n}{\frac{1}{c} I(Z;Y|P = p)}\left[1 + O(\sqrt{\gamma})\right],
\end{align}
thus asymptotically achieving an optimal scaling of the code length for arbitrary values of $p$.
\end{theorem}

Since the asymptotic code length is optimal regardless of $p$, these asymptotics are also optimal when $p$ is optimized to maximize the mutual information in the fully informed setting.

Finally, although it is hard to estimate the scores of mixed tuples with this decoder, just like in~\cite{oosterwijk14} we expect that the joint decoder score for a tuple is roughly equal to the sum of the $c$ individual simple decoder scores. So a tuple of $c$ users consisting of $k$ colluders and $c - k$ innocent users is expected to have a score roughly a factor $k/c$ smaller than the expected score for the all-guilty tuple. So after computing the scores for all tuples of size $c$, we can get rough estimates of how many guilty users are contained in each tuple, and for instance try to find the set $\mathcal{C}'$ of $c$ users that best matches these estimates. There are several options for post-processing that may improve the accuracy of using this joint decoder, which are left for future work.

%WWWWWWWWWWWWWWWWWWWWWWWWWWWWWWWWWWWWWWWWWWWWWWWWWWWWWWWW

\subsection{Fingerprinting attacks}
\label{sec:dec-joint-fp}

Using Theorem~\ref{thm:dec-joint-informed} we can obtain exact expressions for $\ell$ in terms of $\vec{\theta}, p, c, n, \eps_1, \eps_2$. For the optimal values of $p$ of~\cite[Section III.A]{laarhoven14capacities} we can use Theorem~\ref{thm:dec-joint-asymptotics} to obtain the following expressions. Note again that $\ell(\thmin) \sim \ell(\thall)$. 
\begin{align}
\ell(\thint) &= 2 c^2 \ln n \left[1 + O\left(\sqrt{\gamma} + \frac{1}{c^2}\right)\right], \label{eq:dec-joint-int} \\
\ell(\thall) &= c \log_2 n \left[1 + O\left(\sqrt{\gamma} + \frac{1}{c}\right)\right], \\
\ell(\thmaj) &= c \log_2 n \left[1 + O\left(\sqrt{\gamma} + \frac{1}{c}\right)\right], \\
\ell(\thcoi) &= c \log_{5/4} n \left[1 + O\left(\sqrt{\gamma} + \frac{1}{c}\right)\right].
\end{align}
Let us again highlight one resulting decoder in particular; the one for the interleaving attack. In general we can rewrite the ratio inside the logarithm as follows:
\begin{align}
\frac{f_{Z,Y|P}(z,y|p,H_0)}{f_{Z,Y|P}(z,y|p, H_1)} = \frac{f_{Y|Z}(y|z)}{f_{Y|P}(y|p)}.
\end{align}
For the interleaving attack we further have $f_{Y|Z}(1|z) = \frac{z}{c}$ and $f_{Y|P}(1|p) = p$. This leads to the following joint decoder $g$.
\begin{align}
g(z,y,p) = \begin{cases}
\ln(1 - \frac{z}{c}) - \ln(1 - p) & \text{if } y = 0; \\
\ln(\frac{z}{c}) - \ln(p) & \text{if } y = 1.
\end{cases}
\end{align}
This means that the joint scores are purely based on the similarities between the tuple tally $z$ and the expected tuple tally $cp$ for each position $i$. If a tuple's tally $z$ is larger than the expected tally $cp$, then the score is positive if $y = 1$ and negative otherwise, while if $z$ is smaller than $cp$, then the score is positive if $y = 0$ and negative otherwise. For innocent tuples this leads to an expected score of roughly $0$, while for the guilty tuple this leads to a high (positive) expected score.

%WWWWWWWWWWWWWWWWWWWWWWWWWWWWWWWWWWWWWWWWWWWWWWWWWWWWWWWW

\subsection{Group testing models}
\label{sec:dec-joint-gt}

Similar to the above, for group testing models we can also use Theorem~\ref{thm:dec-joint-informed} to obtain exact expressions for $\ell$ in terms of $\vec{\theta}, p, c, n, \eps_1, \eps_2$ with provable error bounds. For the optimal values of $p$ of~\cite[Section III.B]{laarhoven14capacities} we can use Theorem~\ref{thm:dec-joint-asymptotics} to obtain the following refined expressions.
\begin{align}
\ell(\thall) &= c \log_2 n \left[1 + O\left(\sqrt{\gamma} + \frac{1}{c}\right)\right], \\
\ell(\thadd) &= \frac{c \log_2 n}{1 - \frac{1}{2} h(r) + O(r^2)} \left[1 + O\left(\sqrt{\gamma} + \frac{1}{c}\right)\right], \\
\ell(\thdil) &= \frac{c \log_2 n}{1 - \frac{\ln 2}{2} h(r) + O(r^2)} \left[1 + O\left(\sqrt{\gamma} + \frac{1}{c}\right)\right].
\end{align}
Note that as discussed in Theorem~\ref{thm:dec-joint-deterministic} the score function for the classical model is equivalent to simply checking whether some subset of $c$ items matches the test results, i.e.\ whether these would indeed have been the test results, had this subset been the set of defectives. With high probability, only the correct set of defectives passes this test.

%WWWWWWWWWWWWWWWWWWWWWWWWWWWWWWWWWWWWWWWWWWWWWWWWWWWWWWWW%WWWWWWWWWWWWWWWWWWWWWWWWWWWWWWWWWWWWWWWWWWWWWWWWWWWWWWWW

\section{Joint universal decoding}
\label{sec:dec-joint-universal}

Let us now again consider the more common setting in fingerprinting where the attack strategy $\vec{\theta}$ is assumed unknown to the distributor. With the results for simple decoding in mind, and knowing that the interleaving attack is also the asymptotically optimal pirate attack in the joint fingerprinting game, we again turn our attention to the decoder designed against the interleaving attack.

%WWWWWWWWWWWWWWWWWWWWWWWWWWWWWWWWWWWWWWWWWWWWWWWWWWWWWWWW

\subsection{The joint interleaving decoder, revisited}
\label{sec:dec-joint-universal-intro}

Similar to the setting of simple decoding, explicitly proving that this decoder achieves the uninformed capacity is not so easy. However, through a series of reductions we can prove that this decoder is asymptotically capacity-achieving, for certain parameters $\ell$ and $\eta$.

\begin{theorem} \label{thm:dec-joint-universal}
The joint log-likelihood decoder designed against the interleaving attack, using the score function $g$ defined by
\begin{align}
g(z,y,p) = \begin{cases}
\ln(1 - \frac{z}{c}) - \ln(1 - p) & \text{if } y = 0; \\
\ln(\frac{z}{c}) - \ln(p) & \text{if } y = 1.
\end{cases} \label{eq:dec-joint-g-universal}
\end{align}
and the arcsine distribution encoder $f_P^*$ of~\eqref{eq:arcsine} together asymptotically achieve the joint capacity for the uninformed fingerprinting game.
\end{theorem}

\begin{IEEEproof}
First, the simple uninformed capacity is asymptotically equivalent to the joint uninformed capacity, which follows from results of Huang and Moulin~\cite{huang12} and Oosterwijk et al.~\cite{oosterwijk13b} (and Section~\ref{sec:dec-simple}). This means that the simple universal decoder of Theorem~\ref{thm:dec-simple-universal} already asymptotically achieves the joint capacity. We will prove that asymptotically, the proposed universal joint decoder is equivalent to the universal simple decoder of Section~\ref{sec:dec-simple-universal}, thus also achieving the joint uninformed capacity.

Suppose we have a tuple $T$ of size $c$, and suppose in some segment $i$ there are $z$ users who received a $1$ and $c - z$ users who received a $0$. For now also assume that $p \in [\delta, 1 - \delta]$ for some $\delta > 0$ that does not depend on $c$. In case $y = 0$, the combined simple decoder score of this tuple $T$ (using the simple universal decoder $g$ of Section~\ref{sec:dec-simple-universal}) would be:
\begin{align}
\sum_{j \in T} S_{j,i} = z \cdot g(1,0,p) + (c - z) g(0,0,p) \\
= z \ln\left(1 - \frac{1}{c}\right) + (c - z) \ln\left(1 + \frac{p}{c(1-p)}\right) \\
\sim -\frac{z}{c} + \frac{(c - z)p}{c(1 - p)} = \frac{p - z/c}{1 - p}.
\end{align}
On the other hand, if we look at this tuple's joint score with the joint universal decoder $g$ of~\eqref{eq:dec-joint-g-universal}, we have
\begin{align}
S_{T,i} = g(z,0,p) = \ln\left(\frac{1 - z/c}{1 - p}\right) \\
= \ln\left(1 + \frac{p - z/c}{1 - p}\right) \sim \frac{p - z/c}{1 - p}.
\end{align}
Note that the last step follows from the fact that for large $c$, with overwhelming probability we have $z = cp + O(\sqrt{cp})$ (since $Z$ is binomially distributed with mean $cp$ and variance $cp(1-p)$), in which case $(p - z/c)/(1 - p) = o(1)$. Combining the above, we have that $S_{T,i} \sim \sum_{j \in T} S_{j,i}$. So the joint universal decoder score for a tuple $T$ is asymptotically equivalent to the sum of the simple universal decoder scores for the members in this tuple, if $p \in [\delta, 1 - \delta]$. 

Since as argued before the distribution tails $[0, \delta]$ and $[1 - \delta, 1]$ are negligible for the performance of the scheme for sufficiently small $\delta$, and since the same result holds for $y = 1$, the simple and joint decoders are asymptotically equivalent.
\end{IEEEproof}

Note that for the uninformed setting, the simple and joint capacities are asymptotically equivalent, which allowed us to prove the result. For finite $c$ the joint capacity may be slightly higher than the simple capacity, but the fact that they are asymptotically the same does show that there is not as much to gain with joint decoders as there is with e.g.\ joint group testing decoders, where the joint capacity is asymptotically a factor $\log_2(e) \approx 1.44$ higher than the simple capacity. And since assigning joint scores to all tuples of size $c$ for each position $i$ is computationally very involved (and since the resulting joint scores are very similar to the sum of simple universal scores anyway) a more practical choice seems to be to use the simple universal decoder of Section~\ref{sec:dec-simple-universal} instead.

%WWWWWWWWWWWWWWWWWWWWWWWWWWWWWWWWWWWWWWWWWWWWWWWWWWWWWWWW
%WWWWWWWWWWWWWWWWWWWWWWWWWWWWWWWWWWWWWWWWWWWWWWWWWWWWWWWW

\section{Discussion}
\label{sec:discussion}

Let us now briefly discuss the main results in this paper and their consequences.

\subsection{Simple informed decoding}

For the setting of simple decoders with known collusion channels $\vec{\theta}$, we have shown how to choose the score functions $g$ in the score-based framework, as well as how to choose the threshold $\eta$ and code length $\ell$ to guarantee that certain bounds on the error probabilities are met. With log-likelihood decoders, we showed that these decoders achieve capacity \textit{regardless of $p$ and regardless of $\vec{\theta}$}. This means that no matter which pirate strategy $\vec{\theta}$ and bias $p$ you plug in, this construction will always achieve the capacity corresponding to those choices for $\vec{\theta}$ and $p$. A trivial consequence is that for the optimal values of $p$ derived in~\cite{laarhoven14capacities}, one also achieves the optimal code length for arbitrary $p$.

\subsection{Simple universal decoding}

Since in fingerprinting it is usually more common to assume that $\vec{\theta}$ is unknown, we then turned our attention to the uninformed setting. We showed that the decoder designed against the interleaving attack, with the score function $g$ given by 
  \begin{align}
g(x,y,p) = \begin{cases} 
\ln\left(1 + \frac{p}{c(1-p)}\right) & \text{if } x = y = 0; \\
\ln\left(1 - \frac{1}{c}\right) & \text{if } x \neq y; \\
\ln\left(1 + \frac{1 - p}{cp}\right) & \text{if } x = y = 1,
\end{cases}
  \end{align}
is actually a universal decoder, and achieves the uninformed fingerprinting capacity. We also showed how the proposed universal decoder is very similar to both Oosterwijk et al.'s decoder $h$ and an approximation of Moulin's empirical mutual information decoder $m$ for $0 \ll p \ll 1$ by
\begin{align}
c \cdot g(x,y,p) \sim h(x,y,p) \sim n \cdot m(x,y,p),
\end{align}
and we highlighted the differences between these decoders for $p \approx 0,1$. We argued that the proposed decoder $g$ is the most natural choice for a universal decoder (motivated from the Neyman-Pearson lemma), and that it has some practical advantages compared to Oosterwijk et al.'s decoder $h$, such as finally being able to get rid of the cut-offs $\delta$ on the density function $f_P$.

\subsection{Joint informed decoding}

In Sections~\ref{sec:dec-joint} and \ref{sec:dec-joint-universal} we then turned our attention to joint decoders, which have the potential to significantly decrease the required code length $\ell$ at the cost of a higher decoding complexity. We considered a natural generalization of the simple decoders to joint decoders, and argued that our choice for the joint score functions $g$ seems to be optimal. There are still some gaps to fill here, since we were not able to prove how scores of mixed tuples (tuples containing some innocent and some guilty users) behave, and whether their scores also stay below $\eta$ with high probability. On the other hand, for deterministic attacks it is quite easy to analyze the behavior of these decoders, and for arbitrary attacks we did show that the code lengths have the optimal asymptotic scaling.

\subsection{Joint universal decoding}

Finally, for the uninformed setting with joint decoders, we proved that the joint interleaving decoder achieves the joint uninformed capacity. Since the joint uninformed capacity is asymptotically the same as the simple uninformed capacity, and since joint decoding generally has a much higher computational complexity than simple decoding, this decoder may not be as practical as the proposed simple universal decoder.

%%WWWWWWWWWWWWWWWWWWWWWWWWWWWWWWWWWWWWWWWWWWWWWWWWWWWWWWWW
%%WWWWWWWWWWWWWWWWWWWWWWWWWWWWWWWWWWWWWWWWWWWWWWWWWWWWWWWW

\section{Open problems}
\label{sec:openproblems}

Finally, let us finish by mentioning some open problems which are left for future work.

%WWWWWWWWWWWWWWWWWWWWWWWWWWWWWWWWWWWWWWWWWWWWWWWWWWWWWWWW

\subsection{Analyzing the simple universal decoder}

While in Section~\ref{sec:dec-simple-universal} we showed that the new simple universal decoder achieves capacity and how one should roughly choose the code length $\ell$ and threshold $\eta$, we did not provide any provable bounds on the error probabilities for the uninformed setting. For earlier versions of Tardos' scheme various papers analyzed such provable bounds~\cite{blayer08, ibrahimi13, laarhoven14dcc, oosterwijk13b, skoric08} and a similar analysis could be done for the log-likelihood decoder designed against the interleaving attack. Perhaps such an analysis may once and for all establish the best way to choose the scheme parameters in universal fingerprinting.

%WWWWWWWWWWWWWWWWWWWWWWWWWWWWWWWWWWWWWWWWWWWWWWWWWWWWWWWW

\subsection{Dynamic fingerprinting and adaptive group testing}

Although this paper focused on applications to the `static' fingerprinting game, in some settings the feedback $Y$ may be obtained in real-time. For instance, in pay-tv pirates may try to duplicate a fingerprinted broadcast in real time, while in group testing it may sometimes be possible to do several rounds of group testing sequentially. The construction of~\cite{laarhoven13tit} can trivially be applied to the decoders in this paper as well to build efficient dynamic fingerprinting schemes with the same asymptotics for the code lengths, but where (i) the order terms in $\ell$ are significantly smaller; (ii) one can provably catch \textit{all} pirates regardless of the (asymmetric) pirate strategy; and (iii) one does not necessarily need to know (a good estimate of) $c$ in advance~\cite[Section~V]{laarhoven13tit}. An important open problem remains to determine the dynamic uninformed fingerprinting capacity, which may prove or disprove that the construction of~\cite{laarhoven13tit} combined with the universal decoder $g$ of Section~\ref{sec:dec-simple-universal} is asymptotically optimal.

%WWWWWWWWWWWWWWWWWWWWWWWWWWWWWWWWWWWWWWWWWWWWWWWWWWWWWWWW

\subsection{Non-binary codes in fingerprinting}

In this work we focused on the binary case of $q = 2$ different symbols for generating the code $\mathcal{X}$, but in (universal) fingerprinting it may be advantageous to work with larger alphabet sizes $q > 2$, since the code length decreases linearly with $q$~\cite{boesten11, huang12b}. This generalization to $q$-ary alphabets was considered in e.g.~\cite{boesten11, huang12b, oosterwijk13b, oosterwijk14, skoric08}. For the results in this paper we did not really use that we were working with a binary alphabet, so it seems a straightforward exercise to prove that the $q$-ary versions of the log-likelihood decoders also achieve the $q$-ary capacities. A harder problem seems to be to actually compute these capacities in the various informed settings, since the maximization problem involved in computing these capacities then transforms from a one-dimensional optimization problem to a $(q - 1)$-dimensional optimization problem.

%WWWWWWWWWWWWWWWWWWWWWWWWWWWWWWWWWWWWWWWWWWWWWWWWWWWWWWWW
%WWWWWWWWWWWWWWWWWWWWWWWWWWWWWWWWWWWWWWWWWWWWWWWWWWWWWWWW

\section*{Acknowledgments} 

The author is grateful to Pierre Moulin for his comments during the author's visit to Urbana-Champaign that lead to the study of the empirical mutual information decoder of Theorem~\ref{thm:dec-simple-moulin} and that eventually inspired work on this paper. The author further thanks Jeroen Doumen, Teddy Furon, Jan-Jaap Oosterwijk, Boris \v{S}kori\'{c}, and Benne de Weger for valuable discussions and comments regarding earlier versions of this manuscript.

%WWWWWWWWWWWWWWWWWWWWWWWWWWWWWWWWWWWWWWWWWWWWWWWWWWWWWWWW
%WWWWWWWWWWWWWWWWWWWWWWWWWWWWWWWWWWWWWWWWWWWWWWWWWWWWWWWW


\begin{thebibliography}{99}

\bibitem{abbe10}
E.~Abbe and L.~Zheng, ``Linear Universal Decoding for Compound Channels," \emph{IEEE Transactions on Information Theory}, vol.~56, no.~12, pp.~5999--6013, 2010.

\bibitem{amiri09}
E.~Amiri and G.~Tardos, ``High Rate Fingerprinting Codes and the Fingerprinting Capacity," \emph{20th ACM-SIAM Symposium on Discrete Algorithms (SODA)}, pp.~336--345, 2009.

\bibitem{atia12}
G.~K.~Atia and V.~Saligrama, ``Boolean Compressed Sensing and Noisy Group Testing," \emph{IEEE Transactions on Information Theory}, vol.~58, no.~3, pp.~1880--1901, 2012.

\bibitem{blayer08}
O.~Blayer and T.~Tassa, ``Improved Versions of Tardos' Fingerprinting Scheme," \emph{Designs, Codes and Cryptography}, vol.~48, no.~1, pp.~79--103, 2008.

\bibitem{boesten11}
D.~Boesten and B.~\v{S}kori\'{c}, ``Asymptotic Fingerprinting Capacity for Non-Binary Alphabets," \emph{13th Conference on Information Hiding (IH)}, pp.~1--13, 2011.

\bibitem{boneh98}
D.~Boneh and J.~Shaw, ``Collusion-Secure Fingerprinting for Digital Data," \emph{IEEE Transactions on Information Theory}, vol.~44, no.~5, pp.~1897--1905, 1998.

\bibitem{chan11}
C.-L.~Chan, P.~H.~Che, S.~Jaggi, and V.~Saligrama, ``Non-adaptive probabilistic group testing with noisy measurements: Near-optimal bounds with efficient algorithms," \emph{49th Allerton Conference on Communication, Control, and Computing}, pp.~1832--1839, 2011.

\bibitem{chan12}
C.-L.~Chan, S.~Jaggi, V.~Saligrama, and S.~Agnihotri, ``Non-Adaptive Group Testing: Explicit Bounds and Novel Algorithms," \emph{IEEE International Symposium on Information Theory (ISIT)}, pp.~1837--1841, 2012.

% Cheraghchi et al.: Dilution noise
\bibitem{cheraghchi11}
M.~Cheraghchi, A.~Hormati, A.~Karbasi, and M.~Vetterli, ``Group Testing with Probabilistic Tests: Theory, Design and Application," \emph{IEEE Transactions on Information Theory}, vol.~57, no.~10, pp.~7057--7067, 2011.

\bibitem{charpentier09}
A.~Charpentier, F.~Xie, C.~Fontaine, and T.~Furon, ``Expectation Maximization Decoding of Tardos Probabilistic Fingerprinting	Code," \emph{SPIE Proceedings / Media Forensics and Security}, vol.~7254, 2009.

\bibitem{cover06}
T.~M.~Cover and J.~A.~Thomas, \emph{Elements of Information Theory (2nd Edition)}, Wiley Press, 2006.

% Dorfman: First work on group testing
\bibitem{dorfman43}
R.~Dorfman, ``The Detection of Defective Members of Large Populations," \emph{The Annals of Mathematical Statistics}, vol.~14, no.~4, pp.~436--440, 1943.

% Dyachkov and Rykov: Deterministic design with T = O(K^2 log N)
\bibitem{dyachkov82}
A.~G.~D'yachkov and V.~V.~Rykov, ``Bounds on the length of disjunctive codes," \emph{Problemy Peredachi Informatsii}, vol.~18, no.~3, pp.~7--13, 1982.

% Dyachkov et al.: Deterministic lower bound of T = Omega(K^2 log N / log K)
\bibitem{dyachkov89}
A.~G.~D'yachkov, V.~V.~Rykov, and A.~M.~Rashad, ``Superimposed distance codes," \emph{Problems of Control and Information Theory}, vol.~18, no.~4, pp.~237--250, 1989.

\bibitem{fiat01}
A.~Fiat and T.~Tassa, ``Dynamic Traitor Tracing," \emph{Journal of Cryptology}, vol.~14, no.~3, pp.~354--371, 2001.

\bibitem{furon09b}
T.~Furon and L.~P\'{e}rez-Freire, ``EM Decoding of Tardos Traitor Tracing Codes," \emph{ACM Symposium on Multimedia and Security (MM\&Sec)}, pp.~99--106, 2009.

\bibitem{huang12}
Y.-W.~Huang and P.~Moulin, ``On the Saddle-Point Solution and the Large-Coalition Asymptotics of Fingerprinting Games," \emph{IEEE Transactions on Information Forensics and Security}, vol.~7, no.~1, pp.~160--175, 2012.

\bibitem{huang12b}
Y.-W.~Huang and P.~Moulin, ``On Fingerprinting Capacity Games for Arbitrary Alphabets and Their Asymptotics," \emph{IEEE International Symposium on Information Theory (ISIT)}, pp.~2571--2575, 2012.

\bibitem{ibrahimi13}
S.~Ibrahimi, B.~\v{S}kori\'{c}, and J.-J.~Oosterwijk, ``Riding the Saddle Point: Asymptotics of the Capacity-Achieving Simple Decoder for Bias-Based Traitor Tracing," \emph{Cryptology ePrint Archive}, 2013.

\bibitem{laarhoven12wifs}
T.~Laarhoven, J.-J.~Oosterwijk, and J.~Doumen, ``Dynamic Traitor Tracing for Arbitrary Alphabets: Divide and Conquer," \emph{IEEE Workshop on Information Forensics and Security (WIFS)}, pp.~240--245, 2012.

\bibitem{laarhoven13ihmmsec}
T.~Laarhoven and B.~de~Weger, ``Discrete Distributions in the Tardos Scheme, Revisited," \emph{1st ACM Workshop on Information Hiding and Multimedia Security (IH\&MMSec)}, pp.~13--18, 2013.

\bibitem{laarhoven13tit}
T.~Laarhoven, J.~Doumen, P.~Roelse, B.~\v{S}kori\'{c}, and B.~de~Weger, ``Dynamic Tardos Traitor Tracing Schemes," \emph{IEEE Transactions on Information Theory}, vol.~59, no.~7, pp.~4230--4242, 2013.

\bibitem{laarhoven13allerton}
T.~Laarhoven, ``Efficient Probabilistic Group Testing Based on Traitor Tracing," \emph{51st Annual Allerton Conference on Communication, Control and Computing (Allerton)}, 2013.

\bibitem{laarhoven13wifs}
T.~Laarhoven, ``Dynamic Traitor Tracing Schemes, Revisited," \emph{IEEE Workshop on Information Forensics and Security (WIFS)}, pp.~191--196, 2013.

\bibitem{laarhoven14dcc}
T.~Laarhoven and B.~de~Weger, ``Optimal Symmetric Tardos Traitor Tracing Schemes," \emph{Designs, Codes and Cryptography}, vol.~71, no.~1, 2014.

\bibitem{laarhoven14ihmmsec}
T.~Laarhoven, ``Capacities and Capacity-Achieving Decoders for Various Fingerprinting Games," \emph{ACM Workshop on Information Hiding and Multimedia Security (IH\&MMSec)}, 2014. This is a preliminary version of the present paper.

\bibitem{laarhoven14capacities}
T.~Laarhoven, ``Asymptotics of Fingerprinting and Group Testing: Tight Bounds from Channel Capacities," \emph{submitted to IEEE Transactions on Information Theory}, 2014.

\bibitem{meerwald11b}
P.~Meerwald and T.~Furon, ``Group Testing Meets Traitor Tracing," \emph{IEEE International Conference on Acoustics, Speech and Signal Processing (ICASSP)}, pp.~4204--4207, 2011.

\bibitem{meerwald12}
P.~Meerwald and T.~Furon, ``Toward Practical Joint Decoding of Binary Tardos Fingerprinting Codes," \emph{IEEE Transactions on Information Forensics and Security}, vol.~7, no.~4, pp.~1168--1180, 2012.

\bibitem{moulin08}
P.~Moulin, ``Universal Fingerprinting: Capacity and Random-Coding Exponents," \emph{arXiv:0801.3837v3 [cs.IT]}, 2011.

\bibitem{neyman33}
J.~Neyman and E.~S.~Pearson, ``On the Problem of the Most Efficient Tests of Statistical Hypotheses," \emph{Philosophical Transactions of the Royal Society A: Mathematical, Physical and Engineering Sciences}, vol.~231, pp.~289--337, 1933.

\bibitem{nuida09}
K.~Nuida, S.~Fujitsu, M.~Hagiwara, T.~Kitagawa, H.~Watanabe, K.~Ogawa, and H.~Imai, ``An Improvement of Discrete Tardos Fingerprinting Codes," \emph{Designs, Codes and Cryptography}, vol.~52, no.~3, pp.~339--362, 2009.

\bibitem{oosterwijk13b}
J.-J.~Oosterwijk, B.~\v{S}kori\'{c}, and J.~Doumen, ``A Capacity-Achieving Simple Decoder for Bias-Based Traitor Tracing Schemes," \emph{Cryptology ePrint Archive}, 2013.

\bibitem{oosterwijk14}
J.-J.~Oosterwijk, J.~Doumen, and T.~Laarhoven, ``Tuple Decoders for Traitor Tracing Schemes," \emph{SPIE Proceedings}, 2014.

\bibitem{perez09}
L.~P\'{e}rez-Freire and T.~Furon, ``Blind Decoder for Binary Probabilistic Traitor Tracing Codes," \emph{IEEE Workshop on Information Forensics and Security (WIFS)}, pp.~46--50, 2009.

\bibitem{sebo85}
A.~Seb\H{o}, ``On Two Random Search Problems," \emph{Journal of Statistical Planning and Inference}, vol.~11, pp.~23--31, 1985.

\bibitem{sejdinovic10}
D.~Sejdinovic and O.~Johnson, ``Note on Noisy Group Testing: Asymptotic Bounds and Belief Propagation Reconstruction," \emph{48th Allerton Conference on Communication, Control, and Computing (Allerton)}, pp.~998--1003, 2010.

\bibitem{skoric08}
B.~\v{S}kori\'{c}, S.~Katzenbeisser, and M.~U.~Celik, ``Symmetric Tardos Fingerprinting Codes for Arbitrary Alphabet Sizes," \emph{Designs, Codes and Cryptography}, vol.~46, no.~2, pp.~137--166, 2008.

\bibitem{stinson00}
D.~R.~Stinson, T.~van~Trung, and R.~Wei, ``Secure Frameproof Codes, Key Distribution Patterns, Group Testing Algorithms and Related Structures," \emph{Journal of Statistical Planning and Inference}, vol.~86, no.~2, pp.~595--617, 2000.

\bibitem{tardos03}
G.~Tardos, ``Optimal Probabilistic Fingerprint Codes," \emph{35th ACM Symposium on Theory of Computing (STOC)}, pp.~116--125, 2003.

\end{thebibliography}
\end{document}